\documentclass[9pt,twocolumn,twoside]{osajnl}
\journal{pr}

\title{Noisy quantum gyroscope}
\author[1]{Lin Jiao}
\author[1,*]{Jun-Hong An}
\affil[1]{Lanzhou Center for Theoretical Physics, Key Laboratory of Theoretical Physics of Gansu Province, Lanzhou University, Lanzhou 730000, China}
\affil[$^{*}$]{Corresponding author: anjhong@lzu.edu.cn}
\date{\today}

\begin{abstract}
Gyroscope for rotation sensing plays a key role in inertial navigation systems. Developing more precise gyroscopes than the conventional ones bounded by classical shot-noise limit by using quantum resources has attracted much attention. However, existing quantum gyroscope schemes suffer severe deterioration under the influence of decoherence, which is called the no-go theorem of noisy metrology. Here, by using two quantized optical fields as quantum probe, we propose a quantum gyroscope scheme breaking through the constraint of the no-go theorem. Our exact analysis of the non-Markovian noise reveals that both the evolution time as a resource in enhancing the sensitivity and the achieved super-Heisenberg limit in the noiseless case are asymptotically recoverable when each optical field forms a bound state with its environment. The result provides a guideline for realizing high-precision rotation sensing in realistic noisy environments.
\end{abstract}

\begin{document}
\maketitle

\section{Introduction}
High-performance gyroscopes for rotation sensing are of pivotal significance for navigation in many types of air, ground, marine, and space applications. Based on the Sagnac effect, i.e., two counter-propagating waves in a rotating loop accumulate a rotation-dependent phase difference, gyroscopes have been realized in optical \cite{Khial2018,Lai2020,Srivastava2016,PhysRevLett.125.033605,PhysRevResearch.2.032069,Sanders:21} and matter-wave \cite{PhysRevLett.78.760,PhysRevLett.78.2046,PhysRevLett.97.240801,PhysRevLett.114.063002,PhysRevLett.114.140404,PhysRevLett.116.183003,PhysRevA.104.013312,PhysRevLett.124.120403} systems. However, the precision of a purely Sagnac gyroscope, which is proportional to the surface area enclosed by the optical path, is theoretically limited by the classical shot-noise limit (SNL). It dramatically constrains their practical application and further performance improvement. The records for precision and stability of commercial gyroscopes are held by optical gyroscopes \cite{Culshaw_2005,LEFEVRE2014851}. To reduce the noise effect, the practical operation of fiber optical gyroscopes generally modulates the optical signal and measures the ratios of harmonics instead of the phase difference \cite{Zhang2018AO}, where classical SNL model is not widely used. Building a purely Sagnac optical gyroscope beating the SNL from the fundamental principle is highly desired.

Pursuing more precise measurement to physical quantities than the classical SNL by using quantum resources \cite{Giovannetti1330,PhysRevLett.96.010401,a2,RevModPhys.89.035002,RevModPhys.90.035005}, such as squeezing \cite{PhysRevD.23.1693,PhysRevLett.118.140401,PhysRevResearch.1.032024} and entanglement \cite{PhysRevLett.112.103604,Luo620}, quantum metrology supplies a way toward achieving gyroscopes with ultimate sensitivity limits. Based on this idea, many schemes of quantum gyroscopes have been proposed. It was found that the entanglement in N00N states \cite{Fink_2019,DeLeonardis2020}, continuous-variable squeezing \cite{Mehmet:10,doi:10.1063/1.5066028,PhysRevApplied.14.034065}, and optical nonlinearity \cite{PhysRevA.94.013830} can enhance the sensitivity of optical gyroscopes beyond the SNL. A quantum-enhanced sensitivity can also be achieved in matter-wave gyroscopes \cite{doi:10.1116/1.5120348,PhysRevLett.125.100402,doi:10.1063/5.0050235,PhysRevA.95.023608} by using spin squeezing \cite{PhysRevA.46.R6797,PhysRevA.47.5138,PhysRevLett.127.083602} or entanglement. However, quantum gyroscopes are still at the stage of the proof-of-principle study, and their superiority over the conventional ones in the absolute value of sensitivity still has not been exhibited \cite{doi:10.1063/5.0050235,RevModPhys.90.035005}. One key obstacle is that the stability of quantum gyroscope is challenged by the decoherence caused by inevitable noise in microscopic world, which generally makes the quantum resources degraded. It was found that the metrology sensitivity using entanglement \cite{PhysRevLett.102.040403,PhysRevA.80.013825,PhysRevLett.107.083601} and squeezing \cite{PhysRevA.81.033819,PhysRevA.95.053837} exclusively returns to or even becomes worse than the SNL; thus, their quantum superiority completely disappears when the photon loss is considered. This is called the no-go theorem of noisy quantum metrology \cite{PhysRevLett.116.120801,Albarelli2018restoringheisenberg} and is one difficulty to achieve a high-precision quantum gyroscope in practice.

\begin{figure}[tbp]
 \centering
 \includegraphics[width=\columnwidth]{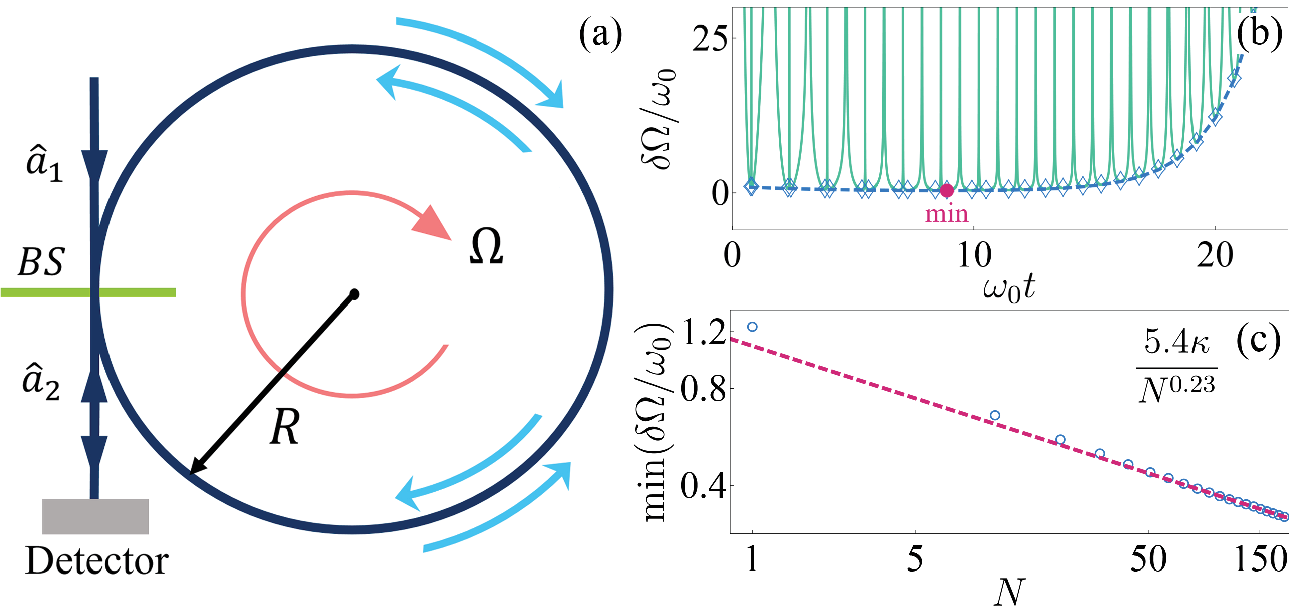}
 \caption{(a) Schematic diagram of quantum gyroscope. (b) Evolution of the error $\delta\Omega_\text{BA}$ (cyan solid line) in the presence of photon loss under the Born-Markovian approximation. The blue dashed line is the local minima of $\delta\Omega_\text{BA}$. The global minimum is marked by the red dot. (c) Numerical fitting reveals that the global minimum scales with the photon number as $\min\delta\Omega_\text{BA}=5.4\kappa N^{-0.23}$. The parameters are $N=100$, $\Omega=\omega_0$, and $\kappa=0.2\omega_0$.	}	\label{scm}
\end{figure}
In this paper, we propose a scheme of quantum gyroscope and discover a mechanism to overcome the constraint of the no-go theorem on our scheme. A super-Heisenberg limit (HL) on the sensitivity is achieved in the ideal case by using two-mode squeezed vacuum state. Our exact analysis on the photon dissipation reveals that the performance of the quantum gyroscope in the dissipative environments intrinsically depends on the energy-spectrum feature of the total system formed by the probe and its environments. The encoding time as resource and the super-HL in the ideal case are asymptotically recovered when each optical field forms a bound state with its environment, which means that the no-go theorem is efficiently avoided. It supplies a guideline to engineer the optimal working condition of our quantum gyroscope in dissipative environments.

\section{Ideal quantum gyroscope scheme}
To measure a physical quantity of certain system, three processes, i.e., the initialization of the quantum probe, the quantity encoding via the probe-system coupling, and the measurement, are generally required. In our quantum gyroscope, we choose two beams of quantized optical fields as the quantum probe. They propagating in opposite directions are input into a 50:50 beam splitter and split into clockwise and counter-clockwise prorogating beams [see Fig. \ref{scm}(a)]. The setup rotates with an angular velocity $\Omega$ about the axis perpendicular to its plane. Thus the two beams accumulate a phase difference $\Delta \theta={\mathcal{N}4\pi k R^2 \Omega/c}$ when they reencounter the beam splitter after $\mathcal{N}$ rounds of propagation in the circular path \cite{scully1997quantum}. Here $k$ is the wave vector, $c$ is the speed of light, and $R$ is the radius of the quantum gyroscope. Remembering the standing-wave condition $kR=n$~($n\in\mathbb{Z}$) of the optical fields propagating along the circular path and defining $\Delta t\equiv \mathcal{N}2\pi R/c$, we have $\Delta\omega\equiv \Delta\theta/\Delta t=2n\Omega$. Therefore, the quantum gyroscope can be equivalently treated as two counter-propagating optical fields with a frequency difference $\Delta\omega$ along the circular path. For concreteness, we choose the basic mode $n=1$. Then the optical fields in the quantum gyroscope can be quantum mechanically described by $(\hbar=1)$ \cite{PhysRevA.95.012326}
\begin{equation}\label{3}
	\hat{H}_{S}=\omega_{0}\sum_{l=1,2}\hat{a}_{l}^{\dagger}\hat{a}_{l}+\Omega(\hat{a}_{1}^{\dagger}\hat{a}_{1}-\hat{a}_{2}^{\dagger}\hat{a}_{2}).
\end{equation}where $\hat{a}_l$ is the annihilation operator of the $l$th field with frequency $\omega_0$. The optical fields couple to the beam splitter twice and output in the state $|\Psi_{out}\rangle =\hat{V}\hat{U}_{0}(\Omega,t)\hat{V}|\Psi_{in}\rangle $, where $\hat{U}_{0}(\Omega,t)=\exp(-i\hat{H}_{S}t)$ is the evolution operator of the fields and $\hat{V}=\exp[i\frac{\pi}{4}(\hat{a}_{1}^{\dagger}\hat{a}_{2}+\hat{a}_{2}^{\dagger}\hat{a}_{1})]$ describes the action of the beam splitter. Thus, the angular velocity $\Omega$ is encoded into the state $|\Psi_{out}\rangle$ of the optical probe via the unitary evolution.

To exhibit the quantum superiority, we employ two-mode squeezed vacuum state as the input state $|\Psi_{in}\rangle=\hat{\mathcal S}|0,0\rangle$, where $\hat{\mathcal S}=\exp[r(\hat{a}_{1}\hat{a}_{2}-\hat{a}_{1}^{\dagger}\hat{a}_{2}^{\dagger})]$ is the squeeze operator, with $r$ being the squeeze parameter. The total photon number of this input state is $N=2\sinh^{2}r$, which is the quantum resource of our scheme. The parity operator $\hat{\Pi}=\exp(i\pi\hat{a}_{1}^{\dagger}\hat{a}_{1})$ is measured at the output port \cite{PhysRevLett.104.103602}. To the output state $|\Psi_{out}\rangle$, we can calculate $\bar{\Pi}\equiv\langle\Psi_{out}|\hat{\Pi}|\Psi_{out}\rangle=[1+N(2+N)\cos^{2}(2\Omega t)]^{-1/2}$ and $\delta\Pi=(1-\bar{\Pi}^{2})^{1/2}$, where $\hat{\Pi}^{2}=1$ has been used. Then the sensitivity of sensing $\Omega$ can be evaluated via the error propagation formula $\delta\Omega=\frac{\delta\Pi}{|\partial_\Omega\bar{\Pi}|}$ as
\begin{equation}
\min\delta\Omega=\big[2t\sqrt{N(2+N)}\big]^{-1},\label{idel}
\end{equation}
when $\Omega t=(2n+1)\pi/4$ with $n\in\mathbb{Z}$. It is remarkable to find that the best sensing error achieved in our scheme is even smaller than the HL $\Delta\Omega\propto(tN)^{-1}$, which reflects the quantum superiority of the used squeezing and measured observable in our scheme. It can be verified that this measurement scheme saturates the Cram\'{e}r-Rao bound governed by quantum Fisher information. We call such a sensitivity surpassing the HL the super-HL \cite{PhysRevX.8.021022,PhysRevLett.126.070503}. It is noted that a phase estimation error smaller than the inverse of the mean photon-number was called the sub-HL in Refs. \cite{PhysRevLett.104.103602,PhysRevLett.108.210404,PhysRevA.88.060101}. The outstanding performance of quantum squeezing has been found in gravitational wave detection \cite{PhysRevLett.124.171101}.

\section{Effects of dissipative environments}
The superiority of quantum sensor is challenged by the decoherence of the quantum probe due to the inevitable interactions with its environment. Depending on whether the probe has energy exchange with the environment or not, the decoherence can be classified into dissipation and dephasing. The main decoherence in our quantum gyroscope is the photon dissipation. The previous works phenomenologically treat the photon dissipation by introducing an imperfect transmission to the beam splitter \cite{PhysRevLett.102.040403,PhysRevA.81.033819,PhysRevA.90.033846,PhysRevA.95.053837,PhysRevLett.108.130402}, which is equivalent to the Born-Markovian approximate description. Such an approximation is convenient, but it might miss important physics. It has been found that the system-environment interplay caused by the inherent non-Markovian nature would induce diverse characters absent in the Born-Markovian approximation \cite{PhysRevA.81.052330,PhysRevLett.109.170402,PhysRevLett.121.220403,RevModPhys.88.021002,LI20181}. To reveal the practical performance of our quantum gyroscope, we, going beyond the Born-Markovian approximation and paying special attention to the non-Markovian effect, investigate the impact of the photon dissipation on the scheme.

We consider that the encoding process is influenced by the photon dissipation, which is caused by the energy exchange between the two optical fields and two independent environments \cite{scully1997quantum}. The Hamiltonian of the total system is
\begin{equation}
\hat{H}=\hat{H}_{S}+\sum_{l=1,2}\sum_{k}[\omega_{k,l}\hat{b}_{k,l}^{\dagger}\hat{b}_{k,l}+g_{k,l}(\hat{a}_{l}^{\dagger}\hat{b}_{k,l}+\text{H.c.})],
\end{equation}
where $\hat{b}_{k,l}$ is the annihilation operator of the $k$th mode with frequency $\omega_{k,l}$ of the environment felt by the $l$th optical field and $g_{k,l}$ is their coupling strength. The coupling is further characterized by the spectral density $J_l(\omega)=\sum_kg_{k,l}^2\delta(\omega-\omega_k)$ in the continuous limit of the environmental frequencies. We consider the Ohmic-family spectral density $J_1(\omega)=J_2(\omega)\equiv J(\omega)=\eta\omega^{s}\omega_{c}^{1-s} e^{-\omega/\omega_{c}}$ for both environments, where $\eta$ is a dimensionless coupling constant, $\omega_{c}$ is a cutoff frequency, and $s$ is an Ohmicity index.
Under the condition that the environments are initially in the vacuum state, we can derive an exact master equation
for the encoding process using the Feynman-Vernon influence functional method \cite{PhysRevA.76.042127},
\begin{equation}	
\dot{\rho}(t)=\sum_{l=1,2}\big\{-i\varpi_l(t)[\hat{a}^\dag_l\hat{a}_l,\rho(t)]+\gamma_{l}(t)\check{\mathcal{D}}_l\rho(t)\big\},\label{maseqt}
\end{equation}
where $\check{\mathcal{D}}_l\cdot=2\hat{a}_{l}\cdot\hat{a}_{l}^{\dagger}-\hat{a}_{l}^{\dagger}\hat{a}_{l}\cdot-\cdot\hat{a}_{l}^{\dagger}\hat{a}_{l}$ is the Lindblad superoperator, $\varpi_l(t)=-\text{Im}[\dot{u}_l(t)/u_l(t)]$ is the renormalized frequency, and $\gamma_l(t)=-\text{Re}[\dot{u}_l(t)/u_l(t)]$ is the dissipation rate. The time-dependent functions $u_l(t)$ satisfy
\begin{equation}\label{eq7}
	\dot{u}_{l}(t)+i\omega_lu_{l}(t)+\int_{0}^{t}f(t-\tau)u_{l}(\tau)d\tau=0,
\end{equation}
under $u_l(0)=1$, where $\omega_{1,2}=\omega_0\pm\Omega$ and $f(x)=\int_{0}^{\infty}J(\omega)e^{-i\omega x}d\omega$ is the environmental correlation function. Equation \eqref{maseqt} indicates that all the non-Markovian effects induced by the environmental backactions have been incorporated into these time-dependent coefficients self-consistently. Solving \eqref{maseqt}, we obtain (see Appendix \ref{smexpv})
\begin{eqnarray}\label{eq8}
	\bar{\Pi}(t)&=&x[4m_{1}(m^*_{2}-m^*_{1}p_{2}^{2})+4m_{2}(m^*_{1}-m^*_{2}p_{1}^{2})\nonumber\\
	&&+(1-p_{1}p_{2})^{2}+16|m_{1}m_{2}|^2]^{-1/2},
\end{eqnarray}
where $x=(\sqrt{A_1A_2}\cosh^2 r)^{-1}$, $m_{l}=\frac{-i u_{l}(t)^{2}\tanh r}{2A_l}$, and $p_{l}=|u_{l}(t)|^{2}(1-A^{-1}_l)$, with $A_l=1-(|u_{l}(t)|^{2}-1)^{2}\tanh^{2}r$. The analytical form of $\delta\Omega$ can then be calculated in a similar manner as the ideal case.

In the special case when the probe-environment coupling is weak and the time scale of $f(t-\tau)$ is smaller than the typical time scale of the probe, we can apply the Born-Markovian approximation in \eqref{eq7} \cite{PhysRevLett.102.040403,PhysRevLett.108.130402}. Their approximate solutions read $u_{l,\text{BA}}(t)=e^{-[\kappa_{l}+i(\omega_{l}+\Delta(\omega_l))]t}$, with $\kappa_{l}=\pi J(\omega_l)$ and $\Delta(\omega_l)=\mathcal{P}\int_{0}^{\infty}\frac{J(\omega)}{\omega_l-\omega}d\omega$. Substituting them into \eqref{eq8} and using the error propagation formula, we obtain (see Appendix \ref{smbmas})
\begin{eqnarray}\label{eq9}
\delta\Omega_\text{BA}(t)=\frac{(2e^{2t\kappa}+NCe^{-2t\kappa})\sqrt{{C}}}{\sqrt{8N}(N+2)t|\sin(4\Omega t)|},
\end{eqnarray}
where $C=4e^{2t\kappa}+N-2+(N+2)\cos(4\Omega t)$. Here we have chosen $\kappa_1=\kappa_2\equiv\kappa$. We plot in Fig. \ref{scm}(b) the evolution of $\delta\Omega_\text{BA}(t)$. It can be found that $\delta\Omega_\text{BA}(t)$ experiences an obvious oscillation with time. However, the best sensitivity manifested by the profile of its local minima tends to be divergent with time. Thus, being in sharp contrast to the ideal case in \eqref{idel}, the superiority of time as a resource in enhancing the precision of the quantum gyroscope disappears. After optimizing the encoding time, we obtain the global minimum $\delta\Omega$ [see the red dot in Fig. \ref{scm}(b)]. The numerical fitting reveals $\min\delta\Omega=5.4\kappa N^{-0.23}$ [see Fig. \ref{scm}(c)], which is even worse than the SNL. Therefore, being consistent with the previous quantum sensing schemes \cite{PhysRevLett.102.040403,PhysRevA.81.033819,PhysRevA.90.033846,PhysRevA.95.053837,PhysRevLett.108.130402}, the photon dissipation under the Born-Markovian approximation makes the quantum advantages of our scheme completely vanish. It is called the no-go theorem of noisy quantum metrology \cite{PhysRevLett.116.120801,Albarelli2018restoringheisenberg} and is the main obstacle to achieve a high-precision quantum sensing in practice.

\begin{figure}[tbp]
\centering
\includegraphics[width=\columnwidth]{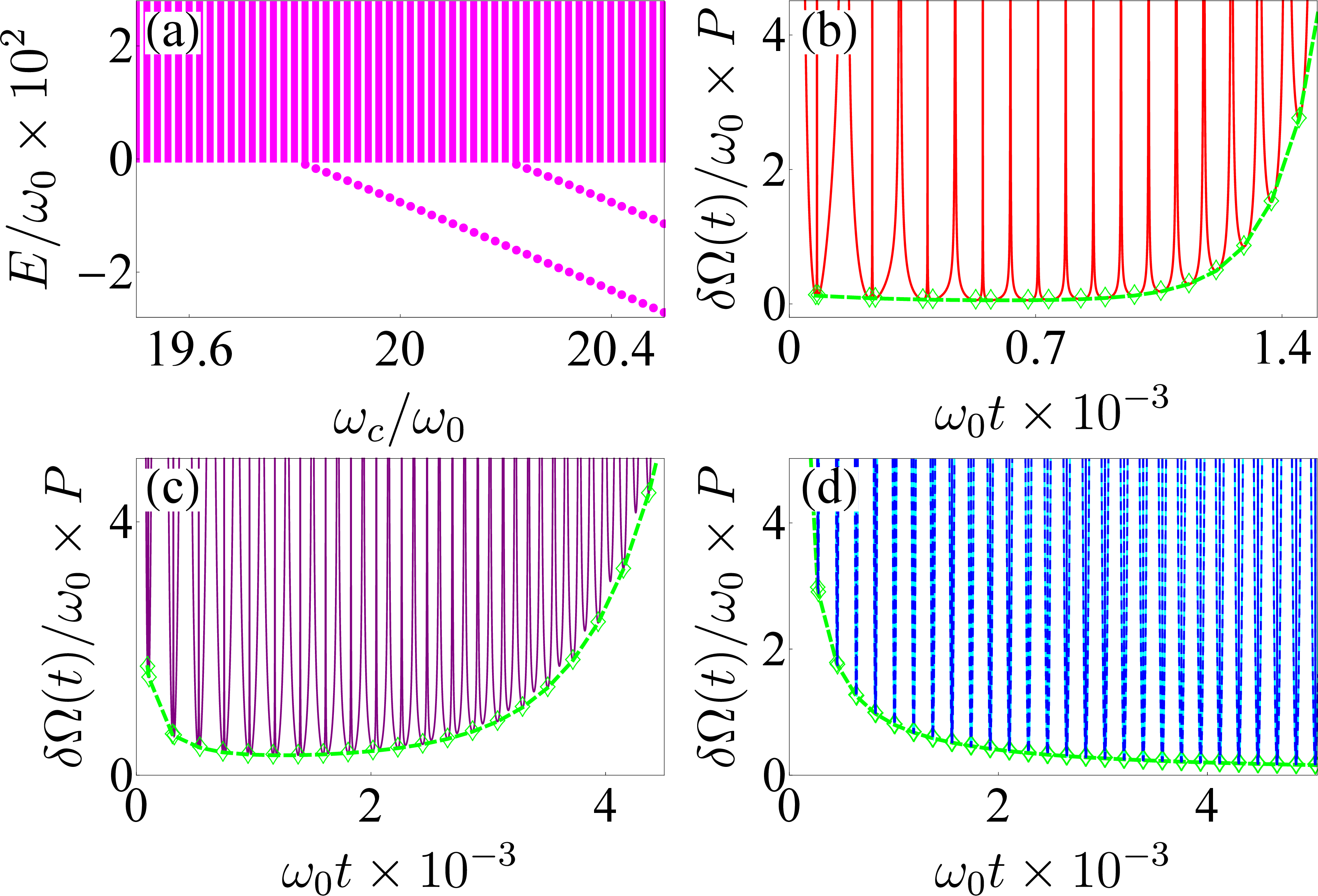}
\caption{(a) Energy spectrum of the total system formed by the two optical fields and their environments. Non-Markovian dynamical evolution of $\delta\Omega(t)$ multiplied by a magnification factor $P$ when $(\omega_c/\omega_0,P)=(2,10^{-1})$ in (b), $(20,10^{-2})$ in (c), and $(25,10^{-3})$ in (d). The blue dashed line in (d) is obtained by numerically solving Eqs. (\ref{eq7}) and the cyan solid line is obtained from the analytical form \eqref{eq12}. We use $s=1$, $\eta=0.05$, $\Omega=10^{-2}\omega_0$, and $N=100$.}\label{1f}
\end{figure}
In the general non-Markovian case, the analytical solution of \eqref{eq7} can be found by the method of Laplace transform, which converts \eqref{eq7} into $\tilde{u}_{l}(z_l)=[z_l+i\omega_l+\int_{0}^{\infty}\frac{J(\omega)d\omega}{z_l+i\omega}d\omega]^{-1}$. Then $u_l(t)$ is obtained by making the inverse Laplace transform to $\tilde{u}_{l}(z_l)$, which can be done by finding its poles from
\begin{equation}\label{eq10}
	Y_l(E_{l})\equiv\omega_l-\int_{0}^{\infty}\frac{J(\omega)}{\omega-E_{l}}d\omega=E_{l},(E_l=iz_l).
\end{equation}
Here, $E_l$ are also the eigenenergies in the single-excitation subspace of the total systems formed by each optical field and its environment. To see this, we expand the eigenstate as $|\Phi_{l}\rangle =(x_{l}\hat{a}_{l}^{\dagger}+\sum_{k}y_{k,l}\hat{b}_{k,l}^{\dagger})|0,\{ 0_{k,l}\} \rangle $. From the stationary Schr\"{o}dinger equation, we have $[E_{l}-(\omega_{0}\pm\Omega)]x_{l}=\sum_{k}g_{k,l}y_{k,l}$ and $y_{k,l}=\hbar g_{k,l}x_{l}/(E_{l}-\hbar\omega_{k,l})$ with $E_l$ being the eigenenergies. The two equations readily result in Eqs. (\ref{eq10}) in the continuous limit of the environmental frequencies. It implies that the dissipation of the optical probe is intrinsically determined by the energy-spectrum character of the probe-environment system in the single-excitation subspace, even though the subspaces with any excitation numbers are involved. Due to $Y_l (E_l)$ are decreasing functions in the regime $E_l < 0$, each of Eqs. (\ref{eq10}) has one isolated root $E_{b,l}$ in this regime provided $Y_l(0)< 0$. While $Y_l(E_l)$ are ill-defined when $E_l > 0$ due to the poles in the integrand, thus they have infinite roots in this regime, which form a continuous energy band. We call the eigenstates of the isolated eigenenergies $E_{b,l}$ bound states \cite{PhysRevA.81.052330}. Making the inverse Laplace transform, we obtain $u_{l}(t)=Z_{l}e^{-iE_{b,l}t}+\int_{0}^{\infty}\Theta(E)e^{-iEt}dE$, where $Z_{l}=[1+\int_{0}^{\infty}\frac{J(\omega)}{(E_{b,l}-\omega)^2}d\omega]^{-1}$ and $\Theta(E)=\frac{J(E)}{[E-\omega_l-\Delta(E)]^{2}+[\pi J(E)]^{2}}$. The integral in $u_l(t)$ is from the energy band and tends to zero in the long-time limit due to the out-of-phase interference. Thus, when the bound state is formed, we have $\lim_{t\rightarrow\infty}u_{l}(t)=Z_{l}e^{-iE_{b,l}t}$, characterizing the suppressed dissipation; otherwise, we have $\lim_{t\rightarrow\infty}u_{l}(t)=0$, meaning a complete dissipation. It can be determined that the bound state is formed for the Ohmic-family spectral density when $\omega_l<\eta\omega_{c}\Gamma(s)$, where $\Gamma(s)$ is the Euler's $\Gamma$ function.

We have three parameter regimes where zero, one, and two bound states are formed, respectively. It is natural to expect that $\delta \Omega$ in the former two regimes is qualitatively consistent with the Born-Markovian approximate result \eqref{eq9} due to the complete dissipation in either two or one optical fields. Focusing on the case in the presence of two bound states and substituting the asymptotic solution $Z_le^{-iE_{b,l}t}$ into \eqref{eq8}, we obtain (see Appendix \ref{smnmdss})
\begin{eqnarray}\label{eq12}
\lim_{t\rightarrow\infty}\delta\Omega(t)=\frac{F\sqrt{2F-4}}{2N(2+N)Z^2_{1}Z^2_{2}t(Z_{1}+Z_{2})|\sin(2Gt)|},~~
\end{eqnarray}
where $F=2+N\sum_lZ_{l}^{2}(2-Z_l^2)+NZ_{1}^{2}Z_{2}^{2}[N+(2+N)\cos(2Gt)]$ with $G=E_{b,1}-E_{b,2}$. We have used $\partial_\Omega E_{b,l}=(-1)^{l-1}Z_l$ derived from \eqref{eq10}. Equation (\ref{eq12}) exhibits a $t^{-1}$-dependence on time, which is as perfect as the ideal result \eqref{idel}. Another finding from \eqref{eq12} is that it tends to the exactly same form as the ideal result \eqref{idel} in the limit $Z_l$ tending to 1. Therefore, the formation of two bound states overcomes the problem of no-go theorem and asymptotically retrieves the ideal sensitivity.

\begin{figure}[tbp]
	\centering
    \includegraphics[width=\columnwidth]{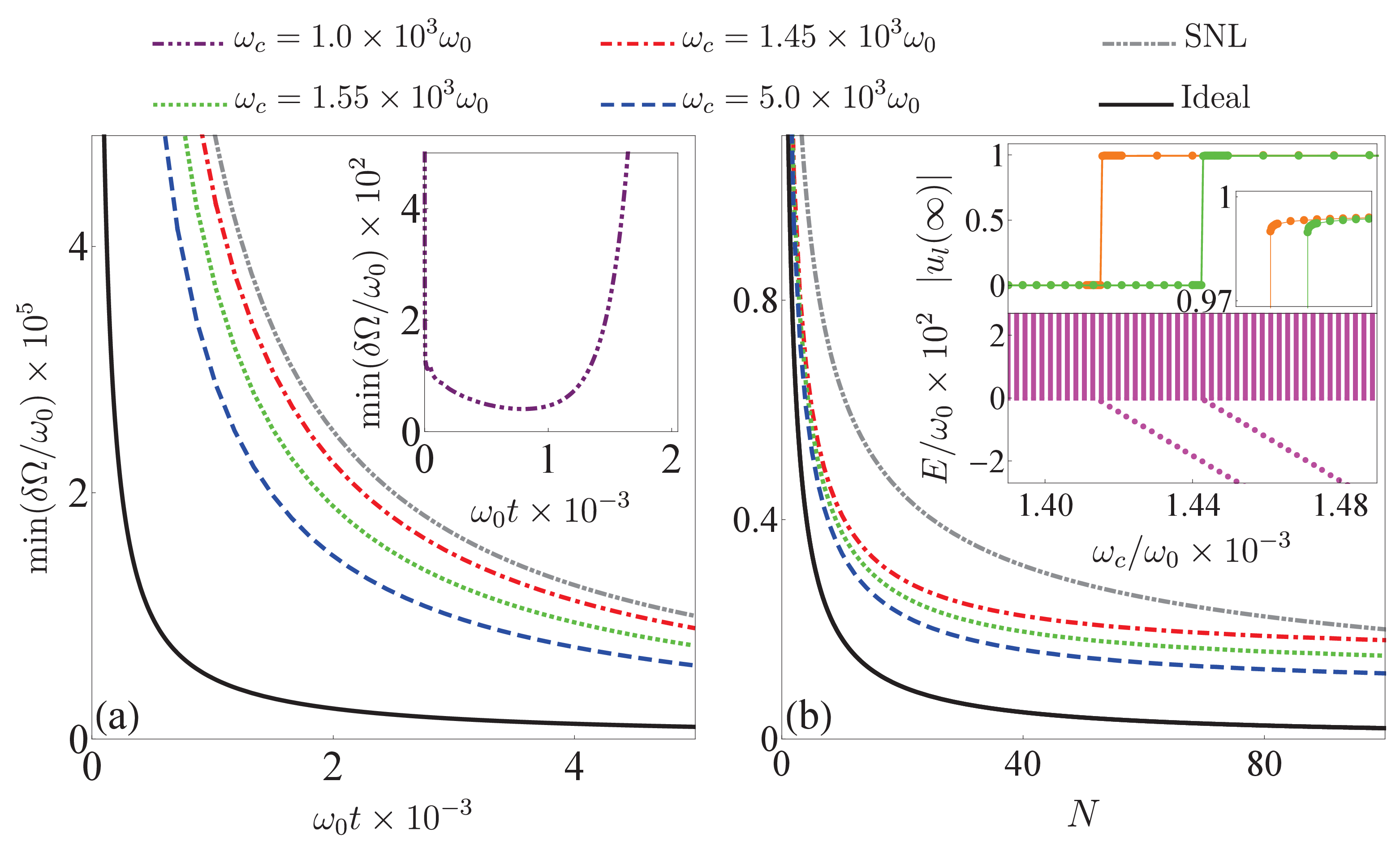}
	\caption{ Local minima of $\delta\Omega(t)$ as a function of time (a) and $N$ when $t =2.5\times 10^4\omega_0^{-1}$ (b) in different $\omega_c$. Steady-state $|u_l(\infty)|$ marked by dots, which match with $Z_l$ depicted by lines, and the energy spectrum are shown in the inset of (b). We use $s=1$, $\eta=7\times 10^{-4}$, $\Omega=10^{-2}\omega_0$, and $N=100$.}
	\label{difomegac}
\end{figure}
\section{Numerical results}
We now numerically verify our general result by choosing the Ohmic spectral density. Figure \ref{1f}(a) shows the energy spectrum of the total system consisting of the optical fields and their environments. It can be seen that the two branches of bound states divide the energy spectrum into three regimes: without bound state when $\omega_c< 19.8\omega_0$, one bound state when $\omega_c\in(19.8,20.2)\omega_0$, and two bound states when $\omega_c>20.2\omega_0$. The result confirms our analytical criterion that the bound states are formed when $\omega_c >\omega_l/[\eta\Gamma(s)]$. Numerically solving \eqref{eq7} and using \eqref{eq8}, we obtain the exact evolution of $\delta\Omega(t)$ in the three regimes. When no or one bound state is formed, the local-minima profile of $\delta\Omega(t)$ tends to divergence in the long-time limit and the quantum superiority of the scheme completely disappears  [see Figs. \ref{1f}(b) and \ref{1f}(c)], which is qualitatively similar to the Born-Markovian result. However, as long as two bound states are formed, the profile of the local minima becomes a decreasing function of the encoding time. The matching of the numerical result with the long-time behavior \eqref{eq12} verifies the validity of the result in \eqref{eq12}. Thus, the encoding time as a resource in sensing $\Omega$ is recovered as perfectly as the ideal case by the formation of the two bound states. Our above result also gives a direct proof on that whether the Born-Markovian approximation is applicable or not depends sensitively on the feature of the energy spectrum of the total probe-environment system. Whenever a bound state is formed in the energy spectrum, the decoherence would be suppressed and the Born-Markovian approximation would no long be valid anymore. This result refreshes our general belief on applicability of the Born-Markovian approximation.

\begin{figure}[tbp]
\centering
\includegraphics[width=\columnwidth]{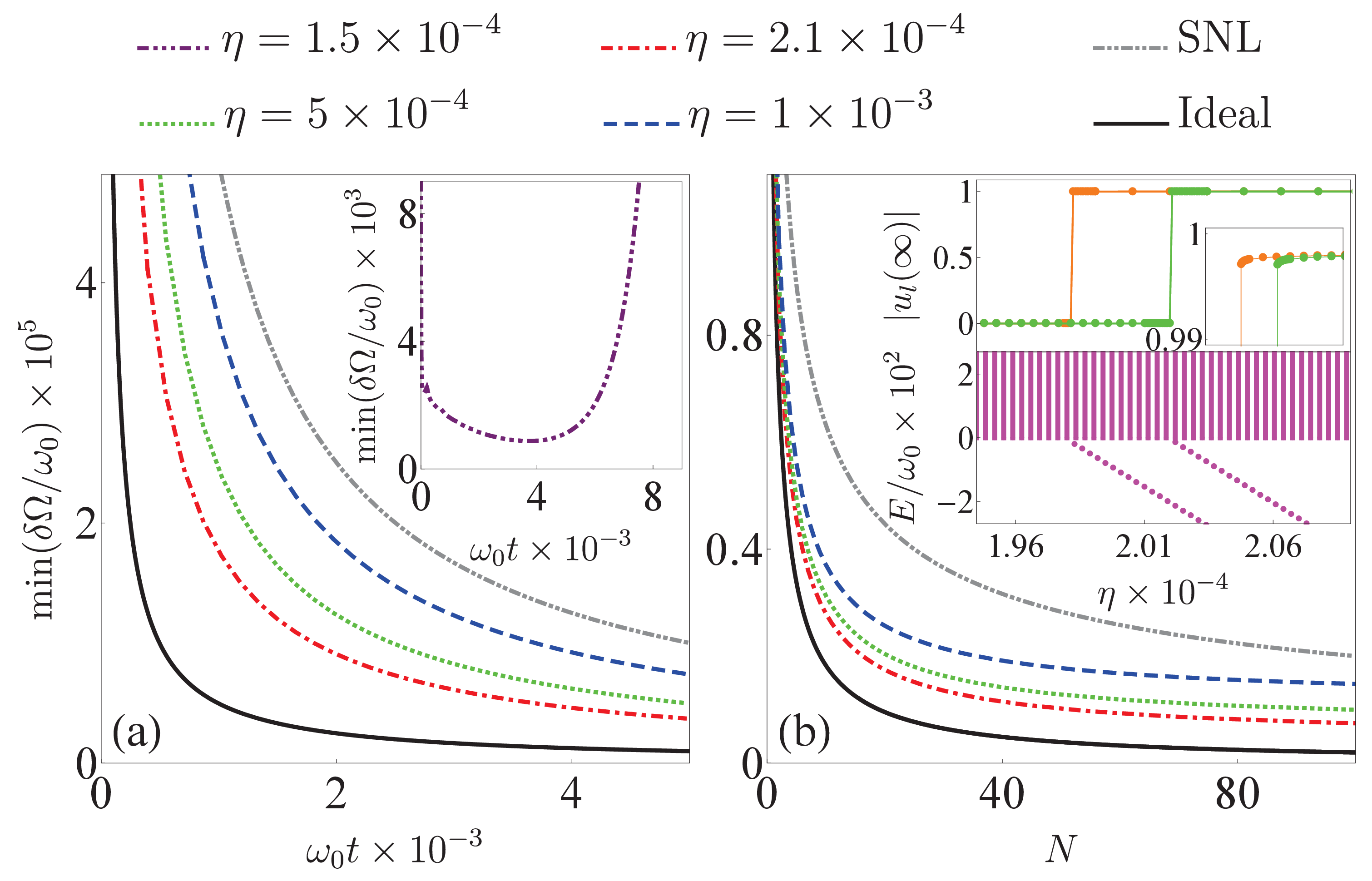}
\caption{Local minima of $\delta\Omega(t)$ as a function of time (a) and $N$ when $t =2.5\times 10^4\omega_0^{-1}$ (b) in different $\eta$. Steady-state $|u_l(\infty)|$ marked by dots, which match with $Z_l$ depicted by lines, and the energy spectrum are shown in the inset of (b). We use $s=1$, $\omega_c=5\times 10^{3}\omega_0$, $\Omega=10^{-2}\omega_0$, and $N=100$.}	\label{difeta}
\end{figure}

Figure \ref{difomegac}(a) shows the evolution of the local minima of $\delta\Omega(t)$ in \eqref{eq12} in different $\omega_c$ when two bound states are formed. The formation of the bound state causes an abrupt increase of the corresponding $|u_l(\infty)|$ from zero to a finite value exactly matching with $Z_l$ [see the inset of Fig. \ref{difomegac}(b)]. It is interesting to find that not only the encoding time as a resource is retrieved, but also the ideal precision is asymptotically recovered. This is double confirmed by the long-time behavior of $\min\delta\Omega(t)$ as a function of the photon number $N$ in Fig. \ref{difomegac}(b). The similar performance is found by changing the coupling constant $\eta$ (see Fig. \ref{difeta}). All the results demonstrate the constructive role played by the two bound states and the non-Markovian effect in retrieving the quantum superiority of our quantum gyroscope. It offers us a guideline to achieve a noise-tolerant rotation sensing by manipulating the formation of the bound states. It is noted that, according to the condition of forming the bound states, we see that what really matters is the relative value $\omega_c/\omega_0$. The equivalent result is achievable by tuning $\omega_0$ for given $\omega_c$ and $\eta$.

\section{Discussion and conclusions}
Our scheme is independent of the form of the spectral density. Although only the Ohmic form is considered, our scheme can be generalized to other spectra. Given the rich way in controlling the spectral density in the setting of quantum reservoir engineering \cite{ER1,Kienzler53}, we deem that our scheme is realizable in state-of-the-art quantum-optical experiments. The Ohmic-family spectral densities for the electromagnetic noise are well cotrolled in circuit QED systems \cite{PhysRevLett.97.016802,Forn-Diaz2017,RevModPhys.86.361}. Actually, the non-Markovian effect has been observed in the linear optical systems \cite{naturephysics1,sciencereport1}. The bound state and its dynamical effect have been observed in circuit QED \cite{nat.phys.1} and ultracold atom \cite{nature5} systems. A squeezing parameter $r\simeq 2.5$, which corresponds to $N\simeq73$, has been realized \cite{Macklin307}. Inspired by these experimental achievement in circuit QED system, we can design a realizable microwave Sagnac interferometer to test our result. We prepare two quantized optical fields propagating in opposite directions in a 1D transmission line, which is coupled via a capacitance or inductance to another wide-band transmission line acting as a structured environment \cite{nat.phys.1}. The squeezed state of the two fields can be generated by the Josephson traveling-wave parametric amplifier \cite{Macklin307}. When the fields reencounter after several rounds of rotation, a phase difference depending on the measured angular velocity is accumulated. The structured environments, on one hand, exert a strong non-Markovian effect on the dynamics of the fields, on the other hand, protect the scheme from the photon loss according to our mechanism.

In summary, we have proposed a quantum gyroscope scheme by using two quantized fields as quantum probe, which achieves a super-HL sensitivity in measuring the angular velocity. However, the photon dissipation under the conventional Born-Markovian approximation forces this sensitivity even being worse than the classical SNL. To overcome this problem, we have presented a mechanism to retrieve the ideal sensitivity by relaxing this approximation. It is found that the ideal sensitivity is asymptotically recoverable when each optical field forms a bound state with its environment, which can be realized by the technique of quantum reservoir engineering. Exhibiting the optimal working condition of quantum gyroscope, our mechanism breaks through the constraint of the no-go theorem of noisy quantum metrology and supplies a guideline in developing high-precision rotation sensing for next-generation inertial navigation systems.

\appendix
\section{Expectation value of parity operator}\label{smexpv}
In this section, we give the derivation of \eqref{eq8}. The Feynman and Vernon's influence-functional theory enables us to derive the evolution of the reduced density matrix of the quantum probe formed by two quantized optical fields exactly. By expressing the forward and backward evolution operators of the density matrix of the probe and the environments as a double path integral in the coherent-state representation and performing the integration over the environmental degrees of freedom, we incorporate all the environmental effects on the probe in a functional integral named influence functional.  The reduced density matrix fully describing the encoding dynamics of the probe is given by \cite{PhysRevA.76.042127}
\begin{eqnarray}\label{rout}
\rho (\bar{\pmb\alpha}_{f},{\pmb\alpha }_{f}^{\prime };t)
&=&\int d\mu ({\pmb\alpha }_{i})d\mu ({\pmb\alpha }%
_{i}^{\prime })\mathcal{J}(\bar{\pmb\alpha}_{f},{\pmb\alpha }%
_{f}^{\prime };t|\bar{\pmb \alpha}_{i},{\pmb\alpha }%
_{i}^{\prime };0)  \notag \\
&&~~~~~~~~~\times \rho (\bar{\pmb\alpha}_{i},{\pmb\alpha }%
_{i}^{\prime };0),
\end{eqnarray}%
where $\rho(\bar{\pmb\alpha}_f,{\pmb\alpha}^{\prime}_f; t) =
\langle\bar{\pmb\alpha}_f|\rho(t)|{\pmb\alpha}^{\prime}_f\rangle$ is the reduced density matrix expressed in coherent-state representation and $\mathcal{J}(\bar{\pmb\alpha}_f, {\pmb\alpha}^{\prime}_f; t|\bar{\pmb\alpha}_i, {\pmb\alpha }^{\prime}_i; 0)$
is the propagating function.  In the derivation of Eq. (\ref{rout}), we have used the coherent-state representation
$|{\pmb\alpha}\rangle = \prod_{l = 1}^2|\alpha_l\rangle$ with $|\alpha_l\rangle = \exp (\alpha _l\hat{a}^{\dagger}_l)|0_l\rangle$,
which are the eigenstates of annihilation operators, i.e., $\hat{a}_l|\alpha_l\rangle = \alpha_l|\alpha_l\rangle$, and obey the
resolution of identity $\int d\mu \left( {\pmb\alpha}\right) |{\pmb\alpha }\rangle \langle \bar{\pmb\alpha }|
= 1$ with $d\mu\left({\pmb\alpha}\right) =
\prod_{l}e^{-\bar{\alpha}_{l}\alpha_{l}}\frac{d\bar{\alpha}
_{l}d\alpha _{l}}{2\pi i}$. $\bar{\pmb\alpha}$ denotes the
complex conjugate of ${\pmb\alpha}$.
The propagating function $\mathcal{J}(\bar{\pmb\alpha}_f, {\pmb\alpha}^{\prime}_f; t|\bar{\pmb\alpha}_i, {\pmb\alpha}^{\prime}_i; 0)$ is expressed as the path integral governed by an effective action, which consists of the free actions of the forward and backward
propagators of the optical probe and the influence functional obtained from the integration of environmental degrees of freedom.  After
evaluation of the path integral, its final form reads
\begin{eqnarray}
& & \mathcal{J}(\bar{\pmb\alpha}_{f},{\pmb\alpha }%
_{f}^{\prime};t|\bar{\pmb\alpha}_{i},{\pmb\alpha }%
_{i}^{\prime };0)=\exp \Big\{\sum_{l=1}^{2}\big[u_l(t)\bar{\alpha}%
_{lf}\alpha _{li}  \notag \\
&&~~~~~+\bar{u}_l(t)\bar{\alpha}_{li}^{\prime }\alpha _{lf}^{\prime
}+[1-\left\vert u_l(t)\right\vert ^2]\bar{\alpha}_{li}^{\prime }\alpha _{li}%
\big]\Big\},  \label{prord}
\end{eqnarray}
where $u_l(t)$ satisfies
\begin{equation}\label{ssmut}
\dot{u}_l(t) + i\omega_l u_l(t) + \int^{t}_0 f(t - \tau)u_l(\tau) = 0
\end{equation}
with $f(x) \equiv \int_0^{\infty}J(\omega)e^{-i\omega x}d\omega$ and $u_l(0)=1$. We have assumed that the spectral density of the two environments are identical.

The input state of the probe is a two-mode squeezed vacuum state
$|\Psi_{in}\rangle = \exp[r(\hat{a}_1\hat{a}_2 - \hat{a}^{\dag}_1\hat{a}^{\dag}_2)]|00\rangle$, where $r$ is the squeezing parameter. After passing the first beam splitter of the quantum optical gyroscope, the state changes into $|\Psi(0)\rangle \equiv\hat{V}|\Psi_{in}\rangle$, with $\hat{V}=\exp[i{\pi\over 4}(\hat{a}_1^\dag\hat{a}_2+\hat{a}_2^\dag\hat{a}_1)]$, which acts as the initial state of the encoding dynamics. In the coherent-state representation, this initial state is given by
\begin{equation}
\rho (\bar{\pmb\alpha}_{i},{\pmb\alpha }_{i}^{\prime };0)=%
\frac{1}{\cosh ^{2}r}\exp [-i\frac{\tanh r}{2}\sum_{l}(\bar{\alpha}_{li}^{2}-\alpha _{li}^{\prime 2})].
\end{equation}
The time-dependent reduced density matrix is obtained by integrating the propagating function over the initial state of Eq. (\ref{rout}). It reads
\begin{equation}
\rho (\bar{\pmb\alpha}_{f},{\pmb\alpha }_{f}^{\prime };t)=x\exp [\sum_{l}(m_{l}\bar{\alpha}_{lf}^{2}+\bar{m}_{l}\alpha _{lf}^{\prime 2}+p_{l}\bar{%
\alpha}_{lf}\alpha _{lf}^{\prime })],
\end{equation}
where $x=(\sqrt{A_1A_2}\cosh^2 r)^{-1}$, $m_{l}=\frac{-i u_{l}(t)^{2}\tanh r}{2A_l}$, and $p_{l}=|u_{l}(t)|^{2}(1-A^{-1}_l)$, with $A_l=1-(|u_{l}(t)|^{2}-1)^{2}\tanh^{2}r$. Remembering $\rho (t)=\int d\mu ( \boldsymbol{\alpha }_{f}) d\mu(\boldsymbol{\alpha }_{f}^{\prime }) \rho (\boldsymbol{\bar{\alpha}}_{f},\boldsymbol{\alpha }_{f}^{\prime };t)|\alpha_{1f},\alpha_{2f}\rangle\langle\bar{\alpha}_{1f}^{\prime},\bar{\alpha}_{2f}^{\prime }|$ and $\rho_{out}=\hat{V}\rho(t)\hat{V}^\dag$, we obtain
\begin{eqnarray}
&&\rho_{out}=\int d\mu \left( {\pmb\alpha }_{f}\right) d\mu\big({\pmb\alpha }_{f}^{\prime }\big) \rho (\bar{\pmb\alpha}_{f},{\pmb\alpha }_{f}^{\prime };t)\nonumber\\
&&\times|{\alpha_{1f}+i\alpha_{2f}\over\sqrt{2}},{\alpha_{2f}+i\alpha_{1f}\over\sqrt{2}}\rangle\langle{\bar{\alpha}_{1f}^{\prime}-i\bar{\alpha}_{2f}^{\prime}\over \sqrt{2}},{\bar{\alpha}_{2f}^{\prime }-i\bar{\alpha}_{2f}^{\prime}\over\sqrt{2}}|.~~~~~
\end{eqnarray}
Then the expectation value $\bar{\Pi}=Tr[\hat{\Pi}\rho_{out}]$ of the parity operator $\hat{\Pi}=\exp(i\pi\hat{a}_1^\dag\hat{a}_1)$ can be calculated as
\begin{eqnarray}
\bar{\Pi}&=&x[4m_{1}(m^*_{2}-m^*_{1}p_{2}^{2})+4m_{2}(m^*_{1}-m^*_{2}p_{1}^{2})\nonumber\\&&
+(1-p_{1}p_{2})^{2}+16|m_{1}m_{2}|^2]^{-1/2},\label{smexppi}
\end{eqnarray}
where $\hat{\Pi}|\alpha,\beta\rangle=|e^{i\pi}\alpha,\beta\rangle$ has been used.
The sensing sensitivity of $\Omega$ is calculated by $\delta\Omega=\frac{\sqrt{1-\bar{\Pi}^{2}}}{|\partial_\Omega\bar{\Pi}|}$.

In the ideal limit, the solution of \eqref{ssmut} reads $u_l(t)=\exp(-i\omega_l t)$ and thus $A_l=1$, $p_l=0$, and $m_l={-ie^{-i2\omega_l t}\tanh r\over 2}$. Then \eqref{smexppi} reduces to
$\bar{\Pi}=[1+N(2+N)\cos^{2}(2\Omega t)]^{-1/2}$ with $N=2\sinh^2 r$.

\begin{figure*}[tbp]
\centering
\includegraphics[width=0.9\textwidth]{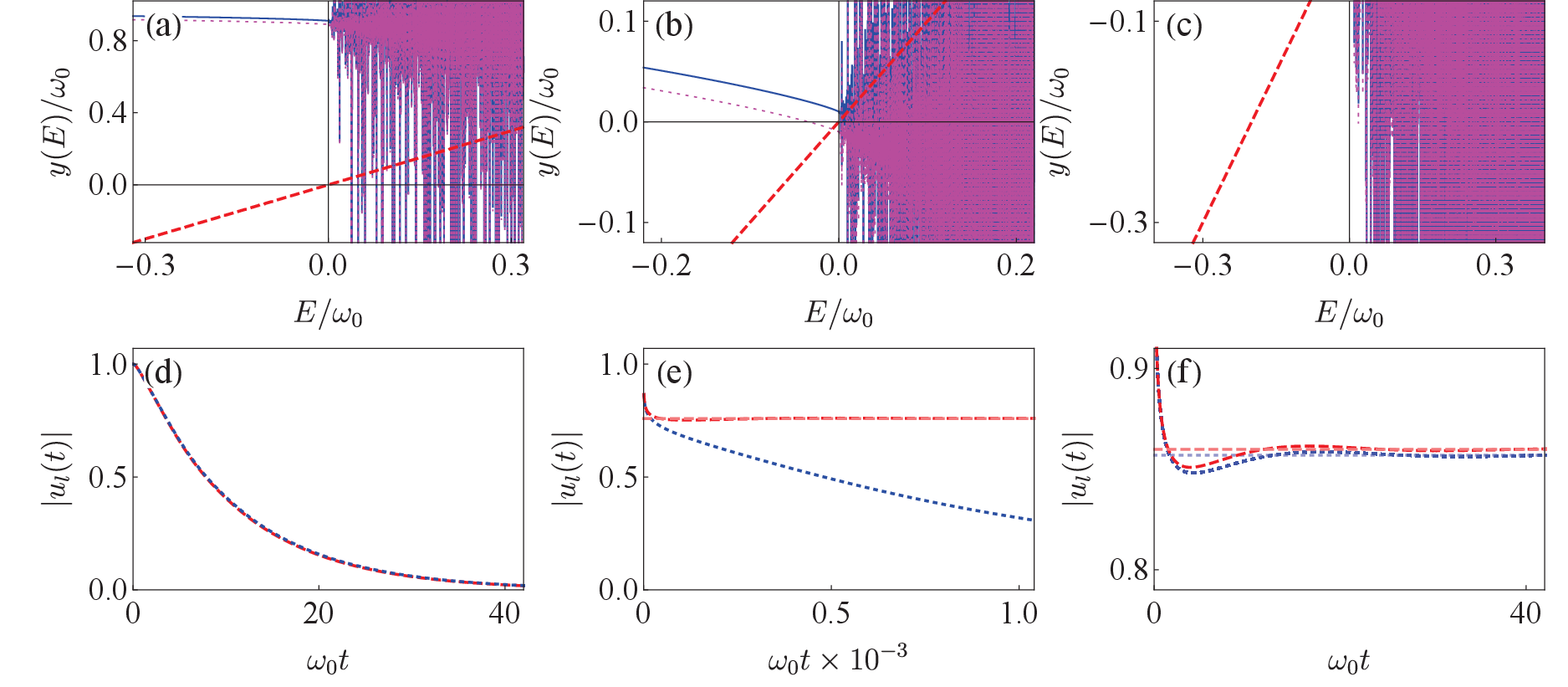}
\caption{(a, b, c) Solution of \eqref{smeq20} determined by the intersectors of two curves of $y(E) = E$ (red dashed lines) and $y(E) = Y_+(E)$ (blue solid lines) or $y(E)=Y_-(E)$ (magenta dotted lines). In the regime $E>0$, both $Y_\pm(E)$ have infinite intersections with $E$, which form a continuous energy band. As long as either $Y_-(0)<0$ or $Y_+(0)<0$, an isolated eigenenergy corresponding to a bound state is formed in the regime $E<0$. (d, e, f) Corresponding behaviors of $|u_+(t)|$ (blue dotted lines) and $|u_-(t)|$ (red dashed lines) determined by numerically solving \eqref{ssmut}. The light-blue dotted and light-red dashed lines in (d) and (e) shows $Z_l$ determined by  \eqref{smzl}. Accompanying the formation of a bound state, the corresponding $|u_l(t)|$ approaches a finite value, which exactly matches with $Z_l$.
The parameters are $s=1$, $\eta=0.05$, $\Omega=10^{-2}\omega_0$, $\omega_c=2\omega_0$ in (a) and (d), $20\omega_0$ in (b) and (e), and $25\omega_0$ in (c) and (f).  } \label{smsbst}
\end{figure*}

\section{Sensitivity under the Born-Markovian approximation}\label{smbmas}
 Defining $u_l(t)=e^{-i\omega_l t}u'_l(t)$, we can rewritten \eqref{ssmut} as
\begin{equation}
\dot{ u}'_l(t)+\int_0^td\tau\int_0^\infty d\omega J(\omega)e^{-i(\omega-\omega_l)(t-\tau)}u'_l(\tau)=0.\label{su}
\end{equation}When the probe-environment coupling is weak and the time scale of the environmental correlation function is much smaller than the one of the probe, we can apply the Born-Markovian approximation to Eq. (\ref{su}) by neglecting the memory effect, i.e., $u'(\tau)\simeq u'(t)$, and extending the upper limit of the integral to infinity, i.e., $\int_0^td\tau\simeq\int_0^\infty d\tau$. The utilization of the identity $\lim_{t\rightarrow\infty}\int_0^t d\tau e^{-i(\omega-\omega_0)(t-\tau)}=\pi\delta(\omega-\omega_0)+i\mathcal{P}{1\over \omega_0-\omega}$, with $\mathcal{P}$ being the Cauchy principal value, results in $u'_{l,\text{MA}}(t)=e^{-[\kappa_l+i\Delta(\omega_l)]t}$, where $\kappa_l=\pi J(\omega_l)$ and $\Delta(\omega_l)=\mathcal{P}\int_0^\infty{J(\omega)\over\omega_l-\omega}d\omega$. We, thus, have the Born-Markovian approximate solution of $u_l(t)$ as $u_{l,\text{MA}}(t)=e^{-[\kappa_l+i(\omega_l+\Delta(\omega_l))]t}$.

Substituting $u_{l,\text{MA}}(t)$ into \eqref{smexppi} and using the error propagation formula, we analytically obtain the sensitivity under the Born-Markovian approximation as
\begin{eqnarray}
\delta\Omega_\text{BA}(t)=\frac{(2e^{2t\kappa}+NCe^{-2t\kappa})\sqrt{{C}}}{\sqrt{8N}(N+2)t|\sin(4\Omega t)|},\label{smmkdomg}
\end{eqnarray}
where $C=4e^{2t\kappa}+N-2+(N+2)\cos(4\Omega t)$. Here we have chosen $\kappa_1=\kappa_2\equiv\kappa$ and neglected the constant $\Delta(\omega_l)$, which is generally renormalized into $\omega_0$. We readily see from \eqref{smmkdomg} that the sensitivity under the Born-Markovian approximation tends to be divergent in the long-time limit.

\section{Sensitivity in the non-Markovian dynamics}\label{smnmdss}

In the non-Markovian case, \eqref{ssmut} can be analytically solvable by the method of Laplace transform, which converts \eqref{ssmut} into $\tilde{u}_{l}(z_l)=[z_l+i\omega_{l}+\int_{0}^{\infty}\frac{J(\omega)d\omega}{z_l+i\omega}d\omega]^{-1}$. Then $u_l(t)$ is obtained by applying the inverse Laplace transform on $\tilde{u}_{l}(z_{l})$, we obtain
\begin{equation}\label{19}
	u_{l}(t)=\frac{1}{2\pi i}\int_{i\sigma+\infty}^{i\sigma-\infty}\frac{e^{-iE_{l}t}}{E_{l}-\omega_{l}+\int_{0}^{\infty}\frac{J(\omega)}{\omega-E_{l}}d\omega}dE.
\end{equation}
where $E_{l}=iz_{l}$ and $\sigma$ is chosen to be larger than all the poles of the integrand. We find the pole of \eqref{19} from
\begin{equation}\label{smeq20}
	Y_l(E_{l})\equiv\omega_{l}-\int_{0}^{\infty}\frac{J(\omega)}{\omega-E_{l}}d\omega=E_{l}.
\end{equation}
It is noted that $E_l$ are also the eigenenergy in the single-excitation subspace of the total systems formed by each optical field and its environment. To see this, we expand the eigenstate as $|\Phi_{l}\rangle =(x_{l}\hat{a}_{l}^{\dagger}+\sum_{k}y_{k,l}\hat{b}_{k,l}^{\dagger})|0,\{ 0_{k,l}\} \rangle $. From the stationary Schr\"{o}dinger equation, we have $[E_{l}-(\omega_{0}\pm\Omega)]x_{l}=\sum_{k}g_{k,l}y_{k,l}$ and $y_{k,l}=\hbar g_{k,l}x_{l}/(E_{l}-\hbar\omega_{k,l})$, with $E_l$ being the eigenenergy. These two equations readily result in \eqref{smeq20} in the continuous limit of the environmental frequencies.
According to the residue theorem, we have
\begin{eqnarray}
	u_{l}(t)=Z_{l}e^{-iE_{b,l}t}+\int_{0}^{\infty}\Theta(E)e^{-iEt}dE.\label{smullt}
\end{eqnarray}
where
\begin{equation}\label{smzl}
Z_{l}=[1+\int_{0}^{\infty}\frac{J(\omega)}{(E_{b,l}-\omega)^2}d\omega]^{-1}
\end{equation} and $\Theta(E)=\frac{J(E)}{[E-\omega_l-\Delta(E)]^{2}+[\pi J(E)]^{2}}$. The first and the second terms of \eqref{smullt} are the residues contributed from the poles of \eqref{smeq20} in the regime $E_l<0$ and $E_l>0$, respectively. It can be found that $Y_l (E_l)$ are decreasing functions in the regime $E_l < 0$, each of \eqref{smeq20} has one isolated root $E_{b,l}$ in this regime provided $Y_l(0)< 0$. While $Y_l(E_l)$ are not well analytic in the regime $E_l > 0$, thus they have infinite roots in this regime, which form a continuous energy
band. We call the eigenstates of the isolated eigenenergies $E_{b,l}$ bound states. We plot in Fig. \ref{smsbst} the solution of \eqref{smeq20} obtained by the graphical
method. It verifies the three typical features on the solutions, i.e., no bound state when $Y_\pm(0)>0$ in Fig. \ref{smsbst}(a), one bound state when $Y_-(0)<0$ but $Y_+(0)>0$ in Fig. \ref{smsbst}(b), and two bound states when $Y_\pm(0)<0$ in Fig. \ref{smsbst}(c).

Due to the out-of-phase interference, the second term in \eqref{smullt} tends to vanish in the long-time limit. Therefore, we have the asymptotical solution of \eqref{smullt} as
\begin{equation}
\lim_{t\rightarrow\infty}u_l(t) =\begin{cases}0 ,& Y_l(0)\geq0\\Z_{l}e^{-iE_{b,l}t}& Y_l(0)<0 \end{cases}, \label{smasym}
\end{equation}
Since the dominated role played by $u_l(t)$ in the encoding dynamics of the quantum probe, the two qualitatively different asymptotical behaviors in \eqref{smasym} manifest the significance of the bound states in determining the photon dissipation.
The former case of \eqref{smullt} characterizes a complete dissipation, while the latter case denotes a suppressed dissipation. Figures \ref{smsbst}(d), \ref{smsbst}(e), and \ref{smsbst}(f) shows the evolution of $|u_\pm|$. This verifies that, as long as the bound state is formed, the corresponding $|u_l|$ tends to a finite value, which exactly matches $Z_l$.

Focusing on the case in the presence of two bound states and substituting the asymptotic solution $\lim_{t\rightarrow\infty}u_l(t)=Z_le^{-iE_{b,l}t}$ into \eqref{smexppi}, we obtain
\begin{eqnarray}
&&	\lim_{t\rightarrow\infty}\delta\Omega(t)=\frac{F\sqrt{2F-4}}{2N(2+N)}\Big|\frac{\partial_\Omega(Z_{1}^{2}+Z_{2}^{2}-1)^2}{4+2N}+Z^2_{1}Z^2_{2}\big[t(Z_{1}\nonumber\\
	&&~~~+Z_{2})\sin(2Gt)-2\partial_\Omega\ln(Z_1Z_{2})\cos^{2}(Gt)\big]\Big|^{-1}  ,\label{smlmdomg}
\end{eqnarray}
where $F=2+N\sum_lZ_{l}^{2}(2-Z_l^2)+NZ_{1}^{2}Z_{2}^{2}[N+(2+N)\cos(2Gt)]$ with $G=E_{b,1}-E_{b,2}$. We have used $\partial_\Omega E_{b,1}=Z_1$ and $\partial_\Omega E_{b,2}=-Z_2$ derived from \eqref{smeq20}. It can be observed that \eqref{smlmdomg} is dominated by the second term in the large-time limit due to the time-independence of its first and the third terms. We, thus, have
\begin{eqnarray}\label{smeq12}
\lim_{t\rightarrow\infty}\delta\Omega(t)\simeq\frac{F\sqrt{2F-4}}{2N(2+N)Z^2_{1}Z^2_{2}t(Z_{1}+Z_{2})|\sin(2Gt)|}. ~~~~~~~
\end{eqnarray}
Equation (\ref{smeq12}) behaves as $t^{-1}$. Therefore, the problem that the sensing error asymptotically tends to be divergent in the Born-Markovian approximation is overcome.

  \begin{figure}[tbp]
  	\centering
  	\includegraphics[width=\columnwidth]{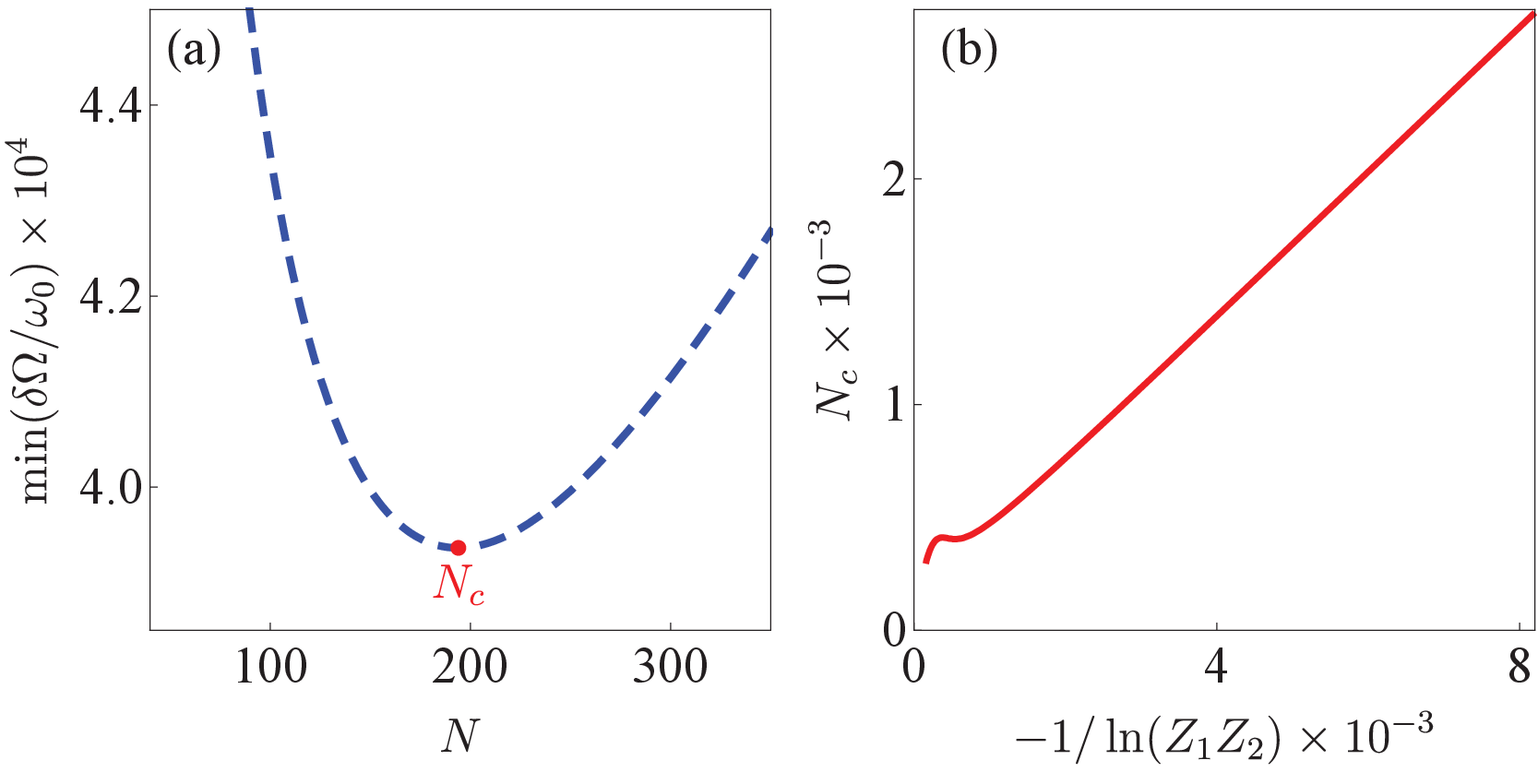}
  	\caption{ (a) Global behavior of the local minima of the steady-state $\delta\Omega(t)$ as a function of $N$. (b) Threshold $N_c$ in (a) as a function $Z_1Z_2$. We use the same parameter values as the ones of the blue solid line in Fig. 4(b) of the main text. }
  	\label{re1}
  \end{figure}

On the one hand, \eqref{smeq12} in the large-$N$ limit behaves as
  \begin{equation}
  \lim_{N\rightarrow\infty}\lim_{t\rightarrow\infty}\delta\Omega(t)\propto N.\label{lna}
  \end{equation}
On the other hand, \eqref{smeq12} in the limit of $Z_l$ tending to 1 behaves as
  \begin{equation}
  \lim_{Z_1,Z_2\rightarrow 1} \lim_{t\rightarrow\infty}\delta\Omega(t)= [2 t\sqrt{N(2+N)}]^{-1},\label{idea}
  \end{equation}
  when $\Omega t=(2n+1)\pi/4$. It is just the ideal sensitivity. Thus the limits of the steady-state $\delta\Omega(t)$ under different orders of $N\rightarrow\infty$ and $Z_l\rightarrow1$ have different scaling relations with $N$. Equations (\ref{lna}) and (\ref{idea}) imply that in the general case $Z_l<1$, the scaling relation of $\lim_{t\rightarrow \infty}\delta\Omega(t)$ with $N$ must have a threshold. Taking the same set of parameter values as the blue dashed line in Fig. \ref{difeta}(b), we plot in Fig. \ref{re1}(a) the global picture of the steady-state $\delta\Omega$ as a function of $N$. There indeed exists a threshold at $N_c\simeq 200$. Below the threshold $N_c$, $\lim_{t\rightarrow\infty}\delta\Omega$ takes the form as \eqref{idea}. Above the threshold $N_c$, it takes the form as \eqref{lna}. Figure \ref{re1}(b) shows the threshold $N_c$ as a function of $Z_1Z_2$. It reveals that the closer $Z_1Z_2$ is to 1, the larger $N_c$ is. In the limit $Z_1Z_2$ tending to 1, $N_c$ tends to an infinity and the scaling behavior \eqref{lna} is completely avoided.

  Experimentally, a squeezing parameter $r\simeq 2.5$, which corresponds to $N\simeq73$, has been realized \cite{Macklin307}. We see from Figs. 3(b) and 4(b) that the threshold is still absent even when $N$ is as large as $100$. Therefore, although we have to face the balance between the bound-state favored restoring superiority and the photon-dissipation caused destruction to the sensitivity, our mechanism still supplies us with a sufficient space to rescue the ideal sensitivity from the noise using the experimentally accessible numbers of quantum resource.

\section*{Funding}
National Natural Science Foundation (Grants No. 11875150, No. 11834005, and No. 12047501).

\section*{Disclosures}
The authors declare no conflicts of interest.

\section*{Data Availability}
 The data that support the plots within this paper and other findings of this paper are available from the corresponding author upon reasonable request.


\begin{thebibliography}{10}
\newcommand{\enquote}[1]{``#1''}

\bibitem{Khial2018}
P.~P. Khial, A.~D. White, and A.~Hajimiri, \enquote{Nanophotonic optical
  gyroscope with reciprocal sensitivity enhancement,}
  \href{http://dx.doi.org/10.1038/s41566-018-0266-5}{{\protect\JournalTitle{Nature
  Photonics}}} \textbf{12}, 671--675 (2018).

\bibitem{Lai2020}
Y.-H. Lai, M.-G. Suh, Y.-K. Lu, B.~Shen, Q.-F. Yang, H.~Wang, J.~Li, S.~H. Lee,
  K.~Y. Yang, and K.~Vahala, \enquote{Earth rotation measured by a chip-scale
  ring laser gyroscope,}
  \href{http://dx.doi.org/10.1038/s41566-020-0588-y}{{\protect\JournalTitle{Nature
  Photonics}}} \textbf{14}, 345--349 (2020).

\bibitem{Srivastava2016}
S.~Srivastava, S.~Rao D.~S., and H.~Nandakumar, \enquote{Novel optical
  gyroscope: proof of principle demonstration and future scope,}
  \href{http://dx.doi.org/10.1038/srep34634}{{\protect\JournalTitle{Sci.
  Rep.}}} \textbf{6}, 34634 (2016).

\bibitem{PhysRevLett.125.033605}
A.~Gebauer, M.~Tercjak, K.~U. Schreiber, H.~Igel, J.~Kodet, U.~Hugentobler,
  J.~Wassermann, F.~Bernauer, C.-J. Lin, S.~Donner, S.~Egdorf, A.~Simonelli,
  and J.-P.~R. Wells, \enquote{Reconstruction of the instantaneous earth
  rotation vector with sub-arcsecond resolution using a large scale ring laser
  array,}
  \href{http://dx.doi.org/10.1103/PhysRevLett.125.033605}{{\protect\JournalTitle{Phys.
  Rev. Lett.}}} \textbf{125}, 033605 (2020).

\bibitem{PhysRevResearch.2.032069}
A.~D.~V. Di~Virgilio, A.~Basti, N.~Beverini, F.~Bosi, G.~Carelli, D.~Ciampini,
  F.~Fuso, U.~Giacomelli, E.~Maccioni, P.~Marsili, A.~Ortolan, A.~Porzio,
  A.~Simonelli, and G.~Terreni, \enquote{Underground {S}agnac gyroscope with
  sub-prad/s rotation rate sensitivity: Toward general relativity tests on
  earth,}
  \href{http://dx.doi.org/10.1103/PhysRevResearch.2.032069}{{\protect\JournalTitle{Phys.
  Rev. Research}}} \textbf{2}, 032069(R) (2020).

\bibitem{Sanders:21}
G.~A. Sanders, A.~A. Taranta, C.~Narayanan, E.~N. Fokoua, S.~A. Mousavi, L.~K.
  Strandjord, M.~Smiciklas, T.~D. Bradley, J.~Hayes, G.~T. Jasion, T.~Qiu,
  W.~Williams, F.~Poletti, and D.~N. Payne, \enquote{Hollow-core resonator
  fiber optic gyroscope using nodeless anti-resonant fiber,}
  \href{http://dx.doi.org/10.1364/OL.410387}{{\protect\JournalTitle{Opt.
  Lett.}}} \textbf{46}, 46--49 (2021).

\bibitem{PhysRevLett.78.760}
A.~Lenef, T.~D. Hammond, E.~T. Smith, M.~S. Chapman, R.~A. Rubenstein, and
  D.~E. Pritchard, \enquote{Rotation sensing with an atom interferometer,}
  \href{http://dx.doi.org/10.1103/PhysRevLett.78.760}{{\protect\JournalTitle{Phys.
  Rev. Lett.}}} \textbf{78}, 760--763 (1997).

\bibitem{PhysRevLett.78.2046}
T.~L. Gustavson, P.~Bouyer, and M.~A. Kasevich, \enquote{Precision rotation
  measurements with an atom interferometer gyroscope,}
  \href{http://dx.doi.org/10.1103/PhysRevLett.78.2046}{{\protect\JournalTitle{Phys.
  Rev. Lett.}}} \textbf{78}, 2046--2049 (1997).

\bibitem{PhysRevLett.97.240801}
D.~S. Durfee, Y.~K. Shaham, and M.~A. Kasevich, \enquote{Long-term stability of
  an area-reversible atom-interferometer {S}agnac gyroscope,}
  \href{http://dx.doi.org/10.1103/PhysRevLett.97.240801}{{\protect\JournalTitle{Phys.
  Rev. Lett.}}} \textbf{97}, 240801 (2006).

\bibitem{PhysRevLett.114.063002}
P.~Berg, S.~Abend, G.~Tackmann, C.~Schubert, E.~Giese, W.~P. Schleich, F.~A.
  Narducci, W.~Ertmer, and E.~M. Rasel, \enquote{Composite-light-pulse
  technique for high-precision atom interferometry,}
  \href{http://dx.doi.org/10.1103/PhysRevLett.114.063002}{{\protect\JournalTitle{Phys.
  Rev. Lett.}}} \textbf{114}, 063002 (2015).

\bibitem{PhysRevLett.114.140404}
R.~Trubko, J.~Greenberg, M.~T.~S. Germaine, M.~D. Gregoire, W.~F. Holmgren,
  I.~Hromada, and A.~D. Cronin, \enquote{Atom interferometer gyroscope with
  spin-dependent phase shifts induced by light near a tune-out wavelength,}
  \href{http://dx.doi.org/10.1103/PhysRevLett.114.140404}{{\protect\JournalTitle{Phys.
  Rev. Lett.}}} \textbf{114}, 140404 (2015).

\bibitem{PhysRevLett.116.183003}
I.~Dutta, D.~Savoie, B.~Fang, B.~Venon, C.~L. Garrido~Alzar, R.~Geiger, and
  A.~Landragin, \enquote{Continuous cold-atom inertial sensor with $1\text{
  }\text{ }\mathrm{nrad}/\mathrm{sec}$ rotation stability,}
  \href{http://dx.doi.org/10.1103/PhysRevLett.116.183003}{{\protect\JournalTitle{Phys.
  Rev. Lett.}}} \textbf{116}, 183003 (2016).

\bibitem{PhysRevA.104.013312}
Y.~Zhao, X.~Yue, F.~Chen, and C.~Huang, \enquote{Extension of the rotation-rate
  measurement range with no sensitivity loss in a cold-atom gyroscope,}
  \href{http://dx.doi.org/10.1103/PhysRevA.104.013312}{{\protect\JournalTitle{Phys.
  Rev. A}}} \textbf{104}, 013312 (2021).

\bibitem{PhysRevLett.124.120403}
E.~R. Moan, R.~A. Horne, T.~Arpornthip, Z.~Luo, A.~J. Fallon, S.~J. Berl, and
  C.~A. Sackett, \enquote{Quantum rotation sensing with dual {S}agnac
  interferometers in an atom-optical waveguide,}
  \href{http://dx.doi.org/10.1103/PhysRevLett.124.120403}{{\protect\JournalTitle{Phys.
  Rev. Lett.}}} \textbf{124}, 120403 (2020).

\bibitem{Culshaw_2005}
B.~Culshaw, \enquote{The optical fibre {S}agnac interferometer: an overview of
  its principles and applications,}
  \href{http://dx.doi.org/10.1088/0957-0233/17/1/r01}{{\protect\JournalTitle{Measurement
  Science and Technology}}} \textbf{17}, R1--R16 (2005).

\bibitem{LEFEVRE2014851}
H.~C. Lef\`{e}vre, \enquote{The fiber-optic gyroscope, a century after
  {S}agnac's experiment: The ultimate rotation-sensing technology?}
  \href{http://dx.doi.org/https://doi.org/10.1016/j.crhy.2014.10.007}{{\protect\JournalTitle{Comptes
  Rendus Physique}}} \textbf{15}, 851--858 (2014).

\bibitem{Zhang2018AO}
H.~Zhang, X.~Chen, X.~Shu, and C.~Liu, \enquote{Fiber optic gyroscope noise
  reduction with fiber ring resonator,}
  \href{http://dx.doi.org/10.1364/AO.57.007391}{{\protect\JournalTitle{Appl.
  Opt.}}} \textbf{57}, 7391--7397 (2018).

\bibitem{Giovannetti1330}
V.~Giovannetti, S.~Lloyd, and L.~Maccone, \enquote{Quantum-enhanced
  measurements: Beating the standard quantum limit,}
  \href{http://dx.doi.org/10.1126/science.1104149}{{\protect\JournalTitle{Science}}}
  \textbf{306}, 1330--1336 (2004).

\bibitem{PhysRevLett.96.010401}
V.~Giovannetti, S.~Lloyd, and L.~Maccone, \enquote{Quantum metrology,}
  \href{http://dx.doi.org/10.1103/PhysRevLett.96.010401}{{\protect\JournalTitle{Phys.
  Rev. Lett.}}} \textbf{96}, 010401 (2006).

\bibitem{a2}
V.~Giovannetti, S.~Lloyd, and L.~Maccone, \enquote{Advances in quantum
  metrology,}
  \href{http://dx.doi.org/10.1038/nphoton.2011.35}{{\protect\JournalTitle{Nat.
  Photonics}}} \textbf{5}, 222--229 (2011).

\bibitem{RevModPhys.89.035002}
C.~L. Degen, F.~Reinhard, and P.~Cappellaro, \enquote{Quantum sensing,}
  \href{http://dx.doi.org/10.1103/RevModPhys.89.035002}{{\protect\JournalTitle{Rev.
  Mod. Phys.}}} \textbf{89}, 035002 (2017).

\bibitem{RevModPhys.90.035005}
L.~Pezz\`e, A.~Smerzi, M.~K. Oberthaler, R.~Schmied, and P.~Treutlein,
  \enquote{Quantum metrology with nonclassical states of atomic ensembles,}
  \href{http://dx.doi.org/10.1103/RevModPhys.90.035005}{{\protect\JournalTitle{Rev.
  Mod. Phys.}}} \textbf{90}, 035005 (2018).

\bibitem{PhysRevD.23.1693}
C.~M. Caves, \enquote{Quantum-mechanical noise in an interferometer,}
  \href{http://dx.doi.org/10.1103/PhysRevD.23.1693}{{\protect\JournalTitle{Phys.
  Rev. D}}} \textbf{23}, 1693--1708 (1981).

\bibitem{PhysRevLett.118.140401}
N.~J. Engelsen, R.~Krishnakumar, O.~Hosten, and M.~A. Kasevich, \enquote{Bell
  correlations in spin-squeezed states of 500 000 atoms,}
  \href{http://dx.doi.org/10.1103/PhysRevLett.118.140401}{{\protect\JournalTitle{Phys.
  Rev. Lett.}}} \textbf{118}, 140401 (2017).

\bibitem{PhysRevResearch.1.032024}
D.~Gatto, P.~Facchi, F.~A. Narducci, and V.~Tamma, \enquote{Distributed quantum
  metrology with a single squeezed-vacuum source,}
  \href{http://dx.doi.org/10.1103/PhysRevResearch.1.032024}{{\protect\JournalTitle{Phys.
  Rev. Research}}} \textbf{1}, 032024(R) (2019).

\bibitem{PhysRevLett.112.103604}
Y.~Israel, S.~Rosen, and Y.~Silberberg, \enquote{Supersensitive polarization
  microscopy using {N}{O}{O}{N} states of light,}
  \href{http://dx.doi.org/10.1103/PhysRevLett.112.103604}{{\protect\JournalTitle{Phys.
  Rev. Lett.}}} \textbf{112}, 103604 (2014).

\bibitem{Luo620}
X.-Y. Luo, Y.-Q. Zou, L.-N. Wu, Q.~Liu, M.-F. Han, M.~K. Tey, and L.~You,
  \enquote{Deterministic entanglement generation from driving through quantum
  phase transitions,}
  \href{http://dx.doi.org/10.1126/science.aag1106}{{\protect\JournalTitle{Science}}}
  \textbf{355}, 620--623 (2017).

\bibitem{Fink_2019}
M.~Fink, F.~Steinlechner, J.~Handsteiner, J.~P. Dowling, T.~Scheidl, and
  R.~Ursin, \enquote{Entanglement-enhanced optical gyroscope,}
  \href{http://dx.doi.org/10.1088/1367-2630/ab1bb2}{{\protect\JournalTitle{New
  Journal of Physics}}} \textbf{21}, 053010 (2019).

\bibitem{DeLeonardis2020}
F.~De~Leonardis, R.~Soref, M.~De~Carlo, and V.~M.~N. Passaro, \enquote{On-chip
  group-{I}{V} {H}eisenberg-limited {S}agnac interferometric gyroscope at room
  temperature,}
  \href{http://dx.doi.org/10.3390/s20123476}{{\protect\JournalTitle{Sensors}}}
  \textbf{20}, 3476 (2020).

\bibitem{Mehmet:10}
M.~Mehmet, T.~Eberle, S.~Steinlechner, H.~Vahlbruch, and R.~Schnabel,
  \enquote{Demonstration of a quantum-enhanced fiber {S}agnac interferometer,}
  \href{http://dx.doi.org/10.1364/OL.35.001665}{{\protect\JournalTitle{Opt.
  Lett.}}} \textbf{35}, 1665--1667 (2010).

\bibitem{doi:10.1063/1.5066028}
K.~Liu, C.~Cai, J.~Li, L.~Ma, H.~Sun, and J.~Gao, \enquote{Squeezing-enhanced
  rotating-angle measurement beyond the quantum limit,}
  \href{http://dx.doi.org/10.1063/1.5066028}{{\protect\JournalTitle{Applied
  Physics Letters}}} \textbf{113}, 261103 (2018).

\bibitem{PhysRevApplied.14.034065}
M.~R. Grace, C.~N. Gagatsos, Q.~Zhuang, and S.~Guha, \enquote{Quantum-enhanced
  fiber-optic gyroscopes using quadrature squeezing and continuous-variable
  entanglement,}
  \href{http://dx.doi.org/10.1103/PhysRevApplied.14.034065}{{\protect\JournalTitle{Phys.
  Rev. Applied}}} \textbf{14}, 034065 (2020).

\bibitem{PhysRevA.94.013830}
A.~Luis, I.~Morales, and A.~Rivas, \enquote{Nonlinear fiber gyroscope for
  quantum metrology,}
  \href{http://dx.doi.org/10.1103/PhysRevA.94.013830}{{\protect\JournalTitle{Phys.
  Rev. A}}} \textbf{94}, 013830 (2016).

\bibitem{doi:10.1116/1.5120348}
C.~L. Garrido~Alzar, \enquote{Compact chip-scale guided cold atom gyrometers
  for inertial navigation: Enabling technologies and design study,}
  \href{http://dx.doi.org/10.1116/1.5120348}{{\protect\JournalTitle{AVS Quantum
  Science}}} \textbf{1}, 014702 (2019).

\bibitem{PhysRevLett.125.100402}
S.~S. Szigeti, S.~P. Nolan, J.~D. Close, and S.~A. Haine,
  \enquote{High-precision quantum-enhanced gravimetry with a {B}ose-{E}instein
  condensate,}
  \href{http://dx.doi.org/10.1103/PhysRevLett.125.100402}{{\protect\JournalTitle{Phys.
  Rev. Lett.}}} \textbf{125}, 100402 (2020).

\bibitem{doi:10.1063/5.0050235}
S.~S. Szigeti, O.~Hosten, and S.~A. Haine, \enquote{Improving cold-atom sensors
  with quantum entanglement: Prospects and challenges,}
  \href{http://dx.doi.org/10.1063/5.0050235}{{\protect\JournalTitle{Applied
  Physics Letters}}} \textbf{118}, 140501 (2021).

\bibitem{PhysRevA.95.023608}
C.~Luo, J.~Huang, X.~Zhang, and C.~Lee, \enquote{{H}eisenberg-limited {S}agnac
  interferometer with multiparticle states,}
  \href{http://dx.doi.org/10.1103/PhysRevA.95.023608}{{\protect\JournalTitle{Phys.
  Rev. A}}} \textbf{95}, 023608 (2017).

\bibitem{PhysRevA.46.R6797}
D.~J. Wineland, J.~J. Bollinger, W.~M. Itano, F.~L. Moore, and D.~J. Heinzen,
  \enquote{Spin squeezing and reduced quantum noise in spectroscopy,}
  \href{http://dx.doi.org/10.1103/PhysRevA.46.R6797}{{\protect\JournalTitle{Phys.
  Rev. A}}} \textbf{46}, R6797--R6800 (1992).

\bibitem{PhysRevA.47.5138}
M.~Kitagawa and M.~Ueda, \enquote{Squeezed spin states,}
  \href{http://dx.doi.org/10.1103/PhysRevA.47.5138}{{\protect\JournalTitle{Phys.
  Rev. A}}} \textbf{47}, 5138--5143 (1993).

\bibitem{PhysRevLett.127.083602}
S.-Y. Bai and J.-H. An, \enquote{Generating stable spin squeezing by
  squeezed-reservoir engineering,}
  \href{http://dx.doi.org/10.1103/PhysRevLett.127.083602}{{\protect\JournalTitle{Phys.
  Rev. Lett.}}} \textbf{127}, 083602 (2021).

\bibitem{PhysRevLett.102.040403}
U.~Dorner, R.~Demkowicz-Dobrzanski, B.~J. Smith, J.~S. Lundeen, W.~Wasilewski,
  K.~Banaszek, and I.~A. Walmsley, \enquote{Optimal quantum phase estimation,}
  \href{http://dx.doi.org/10.1103/PhysRevLett.102.040403}{{\protect\JournalTitle{Phys.
  Rev. Lett.}}} \textbf{102}, 040403 (2009).

\bibitem{PhysRevA.80.013825}
R.~Demkowicz-Dobrzanski, U.~Dorner, B.~J. Smith, J.~S. Lundeen, W.~Wasilewski,
  K.~Banaszek, and I.~A. Walmsley, \enquote{Quantum phase estimation with lossy
  interferometers,}
  \href{http://dx.doi.org/10.1103/PhysRevA.80.013825}{{\protect\JournalTitle{Phys.
  Rev. A}}} \textbf{80}, 013825 (2009).

\bibitem{PhysRevLett.107.083601}
J.~Joo, W.~J. Munro, and T.~P. Spiller, \enquote{Quantum metrology with
  entangled coherent states,}
  \href{http://dx.doi.org/10.1103/PhysRevLett.107.083601}{{\protect\JournalTitle{Phys.
  Rev. Lett.}}} \textbf{107}, 083601 (2011).

\bibitem{PhysRevA.81.033819}
T.~Ono and H.~F. Hofmann, \enquote{Effects of photon losses on phase estimation
  near the heisenberg limit using coherent light and squeezed vacuum,}
  \href{http://dx.doi.org/10.1103/PhysRevA.81.033819}{{\protect\JournalTitle{Phys.
  Rev. A}}} \textbf{81}, 033819 (2010).

\bibitem{PhysRevA.95.053837}
Z.~Huang, K.~R. Motes, P.~M. Anisimov, J.~P. Dowling, and D.~W. Berry,
  \enquote{Adaptive phase estimation with two-mode squeezed vacuum and parity
  measurement,}
  \href{http://dx.doi.org/10.1103/PhysRevA.95.053837}{{\protect\JournalTitle{Phys.
  Rev. A}}} \textbf{95}, 053837 (2017).

\bibitem{PhysRevLett.116.120801}
A.~Smirne, J.~Ko\l{}ody\ifmmode~\acute{n}\else \'{n}\fi{}ski, S.~F. Huelga, and
  R.~Demkowicz-Dobrza\ifmmode~\acute{n}\else \'{n}\fi{}ski, \enquote{Ultimate
  precision limits for noisy frequency estimation,}
  \href{http://dx.doi.org/10.1103/PhysRevLett.116.120801}{{\protect\JournalTitle{Phys.
  Rev. Lett.}}} \textbf{116}, 120801 (2016).

\bibitem{Albarelli2018restoringheisenberg}
F.~Albarelli, M.~A.~C. Rossi, D.~Tamascelli, and M.~G. Genoni,
  \enquote{Restoring {H}eisenberg scaling in noisy quantum metrology by
  monitoring the environment,}
  \href{http://dx.doi.org/10.22331/q-2018-12-03-110}{{\protect\JournalTitle{{Quantum}}}}
  \textbf{2}, 110 (2018).

\bibitem{scully1997quantum}
M.~Scully and M.~Zubairy, \emph{Quantum Optics} (Cambridge University Press,
  Cambridge, 1997).

\bibitem{PhysRevA.95.012326}
P.~Kok, J.~Dunningham, and J.~F. Ralph, \enquote{Role of entanglement in
  calibrating optical quantum gyroscopes,}
  \href{http://dx.doi.org/10.1103/PhysRevA.95.012326}{{\protect\JournalTitle{Phys.
  Rev. A}}} \textbf{95}, 012326 (2017).

\bibitem{PhysRevLett.104.103602}
P.~M. Anisimov, G.~M. Raterman, A.~Chiruvelli, W.~N. Plick, S.~D. Huver,
  H.~Lee, and J.~P. Dowling, \enquote{Quantum metrology with two-mode squeezed
  vacuum: Parity detection beats the heisenberg limit,}
  \href{http://dx.doi.org/10.1103/PhysRevLett.104.103602}{{\protect\JournalTitle{Phys.
  Rev. Lett.}}} \textbf{104}, 103602 (2010).

\bibitem{PhysRevX.8.021022}
M.~M. Rams, P.~Sierant, O.~Dutta, P.~Horodecki, and J.~Zakrzewski, \enquote{At
  the limits of criticality-based quantum metrology: Apparent super-heisenberg
  scaling revisited,}
  \href{http://dx.doi.org/10.1103/PhysRevX.8.021022}{{\protect\JournalTitle{Phys.
  Rev. X}}} \textbf{8}, 021022 (2018).

\bibitem{PhysRevLett.126.070503}
Z.~Hou, Y.~Jin, H.~Chen, J.-F. Tang, C.-J. Huang, H.~Yuan, G.-Y. Xiang, C.-F.
  Li, and G.-C. Guo, \enquote{``super-heisenberg'' and heisenberg scalings
  achieved simultaneously in the estimation of a rotating field,}
  \href{http://dx.doi.org/10.1103/PhysRevLett.126.070503}{{\protect\JournalTitle{Phys.
  Rev. Lett.}}} \textbf{126}, 070503 (2021).

\bibitem{PhysRevLett.108.210404}
V.~Giovannetti and L.~Maccone, \enquote{Sub-heisenberg estimation strategies
  are ineffective,}
  \href{http://dx.doi.org/10.1103/PhysRevLett.108.210404}{{\protect\JournalTitle{Phys.
  Rev. Lett.}}} \textbf{108}, 210404 (2012).

\bibitem{PhysRevA.88.060101}
L.~Pezz\'e, \enquote{Sub-heisenberg phase uncertainties,}
  \href{http://dx.doi.org/10.1103/PhysRevA.88.060101}{{\protect\JournalTitle{Phys.
  Rev. A}}} \textbf{88}, 060101 (2013).

\bibitem{PhysRevLett.124.171101}
Y.~Zhao, N.~Aritomi, E.~Capocasa, M.~Leonardi, M.~Eisenmann, Y.~Guo, E.~Polini,
  A.~Tomura, K.~Arai, Y.~Aso, Y.-C. Huang, R.-K. Lee, H.~L\"uck, O.~Miyakawa,
  P.~Prat, A.~Shoda, M.~Tacca, R.~Takahashi, H.~Vahlbruch, M.~Vardaro, C.-M.
  Wu, M.~Barsuglia, and R.~Flaminio, \enquote{Frequency-dependent squeezed
  vacuum source for broadband quantum noise reduction in advanced
  gravitational-wave detectors,}
  \href{http://dx.doi.org/10.1103/PhysRevLett.124.171101}{{\protect\JournalTitle{Phys.
  Rev. Lett.}}} \textbf{124}, 171101 (2020).

\bibitem{PhysRevA.90.033846}
P.~A. Knott, T.~J. Proctor, K.~Nemoto, J.~A. Dunningham, and W.~J. Munro,
  \enquote{Effect of multimode entanglement on lossy optical quantum
  metrology,}
  \href{http://dx.doi.org/10.1103/PhysRevA.90.033846}{{\protect\JournalTitle{Phys.
  Rev. A}}} \textbf{90}, 033846 (2014).

\bibitem{PhysRevLett.108.130402}
J.~J. Cooper, D.~W. Hallwood, J.~A. Dunningham, and J.~Brand, \enquote{Robust
  quantum enhanced phase estimation in a multimode interferometer,}
  \href{http://dx.doi.org/10.1103/PhysRevLett.108.130402}{{\protect\JournalTitle{Phys.
  Rev. Lett.}}} \textbf{108}, 130402 (2012).

\bibitem{PhysRevA.81.052330}
Q.-J. Tong, J.-H. An, H.-G. Luo, and C.~H. Oh, \enquote{Mechanism of
  entanglement preservation,}
  \href{http://dx.doi.org/10.1103/PhysRevA.81.052330}{{\protect\JournalTitle{Phys.
  Rev. A}}} \textbf{81}, 052330 (2010).

\bibitem{PhysRevLett.109.170402}
W.-M. Zhang, P.-Y. Lo, H.-N. Xiong, M.~W.-Y. Tu, and F.~Nori, \enquote{General
  non-{M}arkovian dynamics of open quantum systems,}
  \href{http://dx.doi.org/10.1103/PhysRevLett.109.170402}{{\protect\JournalTitle{Phys.
  Rev. Lett.}}} \textbf{109}, 170402 (2012).

\bibitem{PhysRevLett.121.220403}
H.-J. Zhu, G.-F. Zhang, L.~Zhuang, and W.-M. Liu, \enquote{Universal
  dissipationless dynamics in gaussian continuous-variable open systems,}
  \href{http://dx.doi.org/10.1103/PhysRevLett.121.220403}{{\protect\JournalTitle{Phys.
  Rev. Lett.}}} \textbf{121}, 220403 (2018).

\bibitem{RevModPhys.88.021002}
H.-P. Breuer, E.-M. Laine, J.~Piilo, and B.~Vacchini, \enquote{Colloquium:
  Non-{M}arkovian dynamics in open quantum systems,}
  \href{http://dx.doi.org/10.1103/RevModPhys.88.021002}{{\protect\JournalTitle{Rev.
  Mod. Phys.}}} \textbf{88}, 021002 (2016).

\bibitem{LI20181}
L.~Li, M.~J. Hall, and H.~M. Wiseman, \enquote{Concepts of quantum
  non-{M}arkovianity: A hierarchy,}
  \href{http://dx.doi.org/https://doi.org/10.1016/j.physrep.2018.07.001}{{\protect\JournalTitle{Physics
  Reports}}} \textbf{759}, 1 -- 51 (2018).

\bibitem{PhysRevA.76.042127}
J.-H. An and W.-M. Zhang, \enquote{Non-{M}arkovian entanglement dynamics of
  noisy continuous-variable quantum channels,}
  \href{http://dx.doi.org/10.1103/PhysRevA.76.042127}{{\protect\JournalTitle{Phys.
  Rev. A}}} \textbf{76}, 042127 (2007).

\bibitem{ER1}
C.~J. Myatt, B.~E. King, Q.~A. Turchette, C.~A. Sackett, D.~Kielpinski, W.~M.
  Itano, C.~Monroe, and D.~J. Wineland, \enquote{Decoherence of quantum
  superpositions through coupling to engineered reservoirs,}
  {\protect\JournalTitle{Nature}} \textbf{403}, 269 (2000).

\bibitem{Kienzler53}
D.~Kienzler, H.-Y. Lo, B.~Keitch, L.~de~Clercq, F.~Leupold, F.~Lindenfelser,
  M.~Marinelli, V.~Negnevitsky, and J.~P. Home, \enquote{Quantum harmonic
  oscillator state synthesis by reservoir engineering,}
  \href{http://dx.doi.org/10.1126/science.1261033}{{\protect\JournalTitle{Science}}}
  \textbf{347}, 53--56 (2015).

\bibitem{PhysRevLett.97.016802}
N.-H. Tong and M.~Vojta, \enquote{Signatures of a noise-induced quantum phase
  transition in a mesoscopic metal ring,}
  \href{http://dx.doi.org/10.1103/PhysRevLett.97.016802}{{\protect\JournalTitle{Phys.
  Rev. Lett.}}} \textbf{97}, 016802 (2006).

\bibitem{Forn-Diaz2017}
P.~Forn-D{\'i}az, J.~J. Garc{\'i}a-Ripoll, B.~Peropadre, J.-L. Orgiazzi, M.~A.
  Yurtalan, R.~Belyansky, C.~M. Wilson, and A.~Lupascu, \enquote{Ultrastrong
  coupling of a single artificial atom to an electromagnetic continuum in the
  nonperturbative regime,}
  \href{http://dx.doi.org/10.1038/nphys3905}{{\protect\JournalTitle{Nature
  Physics}}} \textbf{13}, 39--43 (2017).

\bibitem{RevModPhys.86.361}
E.~Paladino, Y.~M. Galperin, G.~Falci, and B.~L. Altshuler, \enquote{$1/f$
  noise: Implications for solid-state quantum information,}
  \href{http://dx.doi.org/10.1103/RevModPhys.86.361}{{\protect\JournalTitle{Rev.
  Mod. Phys.}}} \textbf{86}, 361--418 (2014).

\bibitem{naturephysics1}
B.-H. Liu, L.~Li, Y.-F. Huang, C.-F. Li, G.-C. Guo, E.-M. Laine, H.-P. Breuer,
  and J.~Piilo, \enquote{Experimental control of the transition from
  {M}arkovian to non-{M}arkovian dynamics of open quantum systems,}
  \href{http://dx.doi.org/10.1038/nphys2085}{{\protect\JournalTitle{Nat.
  Phys.}}} \textbf{7}, 931 (2011).

\bibitem{sciencereport1}
N.~K. Bernardes, A.~Cuevas, A.~Orieux, C.~H. Monken, P.~Mataloni, F.~Sciarrino,
  and M.~F. Santos, \enquote{Experimental observation of weak
  non-{M}arkovianity,}
  \href{http://dx.doi.org/10.1038/srep17520}{{\protect\JournalTitle{Sci.
  Rep.}}} \textbf{5}, 17520 (2015).

\bibitem{nat.phys.1}
Y.~Liu and A.~A. Houck, \enquote{Quantum electrodynamics near a
  photonic bandgap,}
  \href{http://dx.doi.org/10.1038/nphys3834}{{\protect\JournalTitle{Nat.
  Phys.}}} \textbf{13}, 48--52 (2012).

\bibitem{nature5}
L.~Krinner, M.~Stewart, A.~Pazmiño, J.~Kwon, and D.~Schneble,
  \enquote{Spontaneous emission of matter waves from a tunable open quantum
  system,}
  \href{http://dx.doi.org/10.1038/s41586-018-0348-z}{{\protect\JournalTitle{Nature}}}
  \textbf{559}, 589--592 (2018).

\bibitem{Macklin307}
C.~Macklin, K.~O{\textquoteright}Brien, D.~Hover, M.~E. Schwartz,
  V.~Bolkhovsky, X.~Zhang, W.~D. Oliver, and I.~Siddiqi, \enquote{A
  near{\textendash}quantum-limited josephson traveling-wave parametric
  amplifier,}
  \href{http://dx.doi.org/10.1126/science.aaa8525}{{\protect\JournalTitle{Science}}}
  \textbf{350}, 307 (2015).

\end{thebibliography}
\end{document}